\documentclass[12pt]{article}
\usepackage{amsfonts, latexsym}
\renewcommand{\theequation}{\arabic{section}.\arabic{equation}}
\newtheorem{proposition}{Proposition}
\newtheorem{lemma}{Lemma}

\begin{document}

\title{\Huge Effective Bosonic Degrees of Freedom for One-Flavour
Chromodynamics}
\author{
    J. Kijowski \\ Center for Theor. Phys., Polish Academy of Sciences \\
		   al. Lotnik\'ow 32/46, 02 - 668 Warsaw, Poland \\ 
		   \ \\
    G. Rudolph, M. Rudolph \\
		   Institut f\"ur Theoretische Physik, Univ. Leipzig \\
		   Augustusplatz 10/11, 04109 Leipzig, Germany  }            
\maketitle

\begin{abstract}
We apply an earlier formulated programme for quantization of nonabelian
gauge theories to one-flavour chromodynamics. This programme consists
in a complete reformulation of the functional integral in terms of
gauge invariant quantities.  For the model under consideration two
types of gauge invariants occur -- quantities, which are bilinear in
quarks and antiquarks (mesons) and a matrix-valued covector field,
which is bilinear in quarks, antiquarks and their covariant
derivatives. This covector field is linear in the original gauge
potential, and can be, therefore, considered as the gauge potential
``dressed'' in a gauge invariant way with matter. Thus, we get a
complete bosonization of the theory. The strong interaction is
described by a highly non-linear effective action obtained after
integrating out quarks and gluons from the functional integral. All
constructions are done consequently on the quantum level, where quarks
and antiquarks are anticommuting objects.  Our quantization procedure
circumvents the Gribov ambiguity.
\end{abstract}

\section{Introduction}

This paper is a continuation of \cite{KR1} and \cite{KRR}. In
\cite{KR1} we have
proved that the classical Dirac-Maxwell system can be reformulated in a
spin-rotation covariant way in terms of gauge invariant quantities and
in
\cite{KRR} we have shown that it is possible to perform similar constructions 
on the level of the (formal) functional integral of Quantum
Electrodynamics. As a result we obtain a functional integral completely
reformulated in terms of local gauge invariant quantities, which
differs essentially from the effective functional integral obtained via
the Faddeev-Popov procedure \cite{FP}. In particular, it turns out that
standard perturbation techniques, based upon a splitting of the
effective Lagrangian into a free part (Gaussian measure) and an
interaction part (proportional to the bare coupling constant $e$), are
rather not applicable to this functional integral. On the contrary, our
formulation seems to be rather well adapted to investigations of
nonperturbative aspects of QED, for a first contribution of this type
see \cite{KRT}.

In this paper we show that our programme can be also applied to a
nonabelian gauge theory, namely Quantum Chromodynamics with one
flavour. As in the case of QED, we end up with a description in terms
of a set $(j^{ab}, {c_{\mu K}}^L)$ of purely bosonic invariants, where
$j^{ab}$ is built from bilinear combinations of quarks and antiquarks
(mesons) and ${c_{\mu K}}^L$ is a set of complex-valued vector bosons
built from the gauge potential and the quark fields. A naive counting
of the degrees of freedom encoded in these quantities yields the
correct result: The field $j^{ab}$ is Hermitean and carries, therefore,
16 degrees of freedom, whereas ${c_{\mu K}}^L$ is complex-valued and
carries, therefore, 32 degrees of freedom. On the other hand, the
original configuration $\left({{A_{\mu}}_A}^B , \psi^a_A \right)$
carries 32 + 24 = 56 degrees of freedom. Thus, exactly 8 gauge degrees
of freedom have been removed.  The main difficulty in our construction
comes, of course, from the fact that the quark fields are
Grassmann-algebra-valued. Ignoring this for a moment, one can give a
heuristical idea, how the invariant vector bosons ${c_{\mu K}}^L$
arise: They may be considered as built from the gauge potential and the
``phase'' of the matter field, ``gauged away'' in a similar way as
within the unitary gauge fixing procedure for theories of nonabelian
Higgs type. (As a matter of fact, in our construction not only the
``phase'', but the whole matter field enters.) In reality, this
simple-minded gauge fixing philosophy cannot be applied to
Grassmann-algebra-valued objects. Instead of that one has to start (in
some sense) with all invariants one can write down. Next one finds
identities relating these invariants, which however --- due to their
Grassmann character --- cannot be ``solved'' with respect to the
correct number of effective invariants. But we show that there exists a
scheme, which enables us to implement these identities under the
functional integral and to integrate out the original quarks and gauge
bosons. This is the main idea of the present paper. As a result we
obtain a functional integral in terms of the correct number of
effective gauge invariant bosonic quantities. Thus, our procedure
consists in a certain reduction to a sector, where we have mesons
$j^{ab}$, whose interaction is mediated by vector bosons ${c_{\mu
K}}^L$.

We stress that our approach circumvents any gauge fixing procedure ---
and, therefore, also the Gribov problem, see \cite{G} -- \cite{Z}. The
whole theory, including the pure Yang-Mills action is rewritten in
terms of invariants. An important property of the effective theory we
obtain is that it is highly non-linear. This is a consequence of
integrating out quarks and gluons. Thus, as in the case of QED, it is
doubtful whether perturbation techniques can be applied here. The
natural next step will be rather to develop a lattice approximation of
QCD within this formulation.

Finally, we mention that our general programme of reformulating gauge
theories in terms of invariants has been earlier applied to theories
with bosonic matter fields (Higgs models), both for the continuum case
\cite{KR2}, \cite{R} and on the 
lattice \cite{KR3}.

The paper is organized as follows: In Section 2 we introduce basic
notations and define gauge invariant quantities, built from the gauge
potential and the (anticommuting) quark fields. Moreover, we prove some
algebraic identities relating these invariants. In Section 3 we derive
basic identities relating the Lagrangian with these invariants.
Finally, in Section 4 we show how to implement the above mentioned
identities under the functional integral to obtain an effective
functional integral in terms of invariants. The paper is completed by
two Appendices, where we give a review of spin tensor algebraic tools
used in this paper, and present some technical points skipped in the
text.

\setcounter{equation}{0}
\section{Basic Notations and Gauge Invariants}

A field configuration of one-flavour chromodynamics consists of an
SU(3)-gauge potential $\left({{A_{\mu}}_A}^B \right)$ and a
four-component colored quark field $\left(\psi^a_A \right)$, where
$A,B, ...,J = 1,2,3$ are color indices, $a,b, ... = 1,2,\dot 1,\dot 2$
denote bispinor indices and $\mu,\nu, ... = 0,1,2,3$ spacetime indices.

Ordinary spinor indices are denoted by $K,L, ... = 1,2$. The components
of $\left(\psi^a_A \right)$ are anticommuting (Grassmann-algebra
valued) quantities and build up a Grassmann-algebra of (pointwise real)
dimension 24.

The one-flavour chromodynamics Lagrangian is given by
\begin{equation} 
\label{Lag}
{\cal L} = {\cal L}_{gauge} + {\cal L}_{mat},
\end{equation}
with
\begin{eqnarray}
{\cal L}_{gauge} & = & - \frac{1}{8} {{F_{\mu \nu}}_A}^B 
{{F^{\mu \nu}}_B}^A, \\
		     \label{Lgauge}
{\cal L}_{mat}   & = & - m \, \overline{\psi^a_A} \, 
	 \beta_{ab} \, g^{AB} \, \psi^b_B 
       - {\rm Im} \left\{ \overline{\psi^a_A} \, \beta_{ab} \,
	 {(\gamma^{\mu})^b}_c \, g^{AB} \, 
	 (D_{\mu} \psi^c_B) \right\}, \label{Lmat}
\end{eqnarray}
where
\begin{eqnarray}
{{F_{\mu \nu}}_A}^B & = & {\partial}_{\mu} {{A_{\nu}}_A}^B
       - {\partial}_{\nu} {{A_{\mu}}_A}^B +
	 i g \,{{\lbrack A_{\mu},A_{\nu} \rbrack}_A}^B, \label{Fmunu} \\
D_{\mu}\psi^a_A & = & {\partial}_{\mu}\psi^a_A + i g \,
	 {{A_{\mu}}_A}^B \, \psi^a_B \label{Dmu}
\end{eqnarray}
are the field strength and the covariant derivative. In contrast to
standard notation, the bar denotes in this paper complex conjugation,
$g^{AB}$ and $\beta_{ab}$ denote the Hermitian metrics in color and
bispinor space respectively and ${\left(\gamma^{\mu}\right)^b}_c$ are
the Dirac matrices (see Appendix A).  The starting point for
formulating the quantum theory is the formal functional measure
\begin{equation}
\label{FunInt1}
{\cal F} = \int \prod{\rm d}\psi \,
	 \prod{\rm d}\overline{\psi} \, \prod{\rm d}A \,
	 {\rm e}^{i \, S \left[ A,\psi,\overline{\psi} \, \right]},
\end{equation}
where
\begin{equation}
\label{FunInt2}
S[A,\psi,\overline{\psi} \, ] = \int {\rm d}^4 x \,
	 {\cal L} \left[ A,\psi,\overline{\psi} \, \right]
\end{equation}
denotes the physical action.  Here, the integral over the anticommuting
fields $\psi$ and $\overline{\psi}$ is understood in the sense of
Berezin (\cite{Ber1}, see also \cite{Ber2} --
\cite{L}). To calculate the vacuum 
expectation value for some observable, i.~e.~a gauge invariant function 
${\cal O}[A,\psi,\overline{\psi}]$, one has to integrate this
observable with respect to the above measure. The main result of our
paper consists in reformulating the measure ${\cal F}$ in terms of
effective, gauge invariant, degrees of freedom of the theory (see
discussion at the end of Section 4). 

Let us define the following fundamental gauge invariant
Grassmann-algebra valued quantities:
\begin{eqnarray}
{\cal J}^{ab} & := & \overline{\psi^a_A}\, g^{AB} \, \psi^b_B, 
\label{Jab} \\
C_{\mu}^{ab} & := & \overline{\psi^a_A} \, g^{AB} \, (D_{\mu} \psi^b_B)
       - (\, \overline{D_{\mu} \psi^a_A} \, ) \, g^{AB} \, \psi^b_B. 
\label{Cmuab}
\end{eqnarray}
Obviously, ${\cal J}^{ab}$ is a Hermitian field of even (commuting)
type (of rank 2 in the Grassmann-algebra):
\begin{equation}
\overline{{\cal J}^{ab}}
	 = \overline{\overline{\psi^a_A} \, g^{AB} \, \psi^b_B}
	 = \overline{\psi^b_B} \, \overline{g^{AB}} \, \psi^a_A
	 = \overline{\psi^b_B} \, g^{BA} \, \psi^a_A
	 = {\cal J}^{ba},
\end{equation}
whereas the field $C_{\mu}^{ab}$ is anti-Hermitian:
\begin{equation}
\overline{C_{\mu}^{ab}} = - \, C_{\mu}^{ba}.
\end{equation}
We denote
\begin{equation}
\label{J2}
{\cal J}^2 := {\cal J}^{ab} \beta_{acbd} {\cal J}^{cd},
\end{equation}
which is a scalar field of rank 4. Here we introduced 
$$
\beta_{abcd} := \frac{1}{2} \, 
         \overline{(\gamma_{\mu})_{ab}} \, (\gamma^{\mu})_{cd}.
$$ Moreover, for shortness of notation we define the following
auxiliary invariants
\begin{eqnarray}
\label{X1}
{\cal X}^2 & := & 4 \, \beta_{bcef} \, \beta_{ad} \, 
       \left\{ {\cal J}^{ad} {\cal J}^{be} {\cal J}^{cf} 
       + 2 \, {\cal J}^{ae} {\cal J}^{bf} {\cal J}^{cd} \right\} 
\nonumber \\
 & \equiv & 4 \, {\cal J}^{ad} {\cal J}^{be} {\cal J}^{cf} \,
       \left\{ \beta_{bcef} \, \beta_{ad} + 2 \, \beta_{bcde} \, 
       \beta_{af} \right\},
\end{eqnarray}
and
\begin{equation}
\label{B}
B_{\mu}^{ab} := \overline{\psi_A^a} \, g^{AB} \, (D_{\mu} \psi_B^b),
\end{equation}
which obey the identities
\begin{eqnarray}
B_{\mu}^{ab} - \overline{B_{\mu}^{ba}} & = & C_{\mu}^{ab}, \label{Id5} \\
B_{\mu}^{ab} + \overline{B_{\mu}^{ba}} & = & \partial_{\mu} {\cal
J}^{ab}\ ,  \label{Id6}
\end{eqnarray}
and consequently
\begin{eqnarray}
B_{\mu}^{ab} & = & \frac{1}{2} \, ((\partial_{\mu} {\cal J}^{ab}) 
	     + C_{\mu}^{ab}), \label{BtoC1} \\
\overline{B_{\mu}^{ba}} & = & \frac{1}{2} \, 
	 ((\partial_{\mu} {\cal J}^{ab}) - C_{\mu}^{ab}). \label{BtoC2}
\end{eqnarray}

\begin{proposition}
\label{Prop0}
The following identities hold
\begin{eqnarray}
\lefteqn{
{\cal X}^2 \, \left( C_{\mu}^{ab} + (\partial_{\mu} {\cal J}^{ab})
\right) } \nonumber \\
  & = & - 4 \, \left( C_{\mu}^{cb} 
       + (\partial_{\mu}{\cal J}^{cb}) \right) \, \beta_{cf} \, 
	 \left( {\cal J}^2 {\cal J}^{af} 
+ 2 \, \beta_{ghde} \, {\cal J}^{gd} {\cal J}^{hf} {\cal J}^{ae} \right) 
	 \label{Id2} \\
  &  & - 8 \, \left( C_{\mu}^{gb} + (\partial_{\mu}{\cal J}^{gb}) \right) \, 
	 \beta_{cf} \, \beta_{ghde} \,
	 \Big( {\cal J}^{he} {\cal J}^{cf} {\cal J}^{ad} 
       + {\cal J}^{cd} {\cal J}^{hf} {\cal J}^{ae} 
       + {\cal J}^{hd} {\cal J}^{ce} {\cal J}^{af} \Big). \nonumber
\end{eqnarray}
\end{proposition}

{\sl Proof.}
To prove these identities we make use of the following relations:
\begin{eqnarray}
{\overline{\epsilon}}^{ABC} \,{\epsilon}^{DEF} & = &
	 g^{AD} \, g^{BE} \, g^{CF} + g^{AE} \, g^{BF} \, g^{CD} +
	 g^{AF} \, g^{BD} \, g^{CE} \nonumber\\
 &   &   - g^{AF} \, g^{BE} \, g^{CD} - g^{AE} \, g^{BD} \,
	 g^{CF} - g^{AD} \, g^{BF} \, g^{CE}, \label{ee} \\
\epsilon^{ABC} \, g^{EF} & = &
	 \epsilon^{ABF} \, g^{EC} + \epsilon^{BCF} \, g^{EA}
	 + \epsilon^{CAF} \, g^{EB} \label{eg}.
\end{eqnarray}         
Using the symmetry properties of $\beta_{ghde}$ (see Appendix A) we first 
calculate
\begin{eqnarray*}
{\cal X}^2 \, B_{\mu}^{ab} & = & {\cal X}^2 \, g^{AB} \,
\overline{\psi_A^a} \,
				( D_{\mu} \psi_B^b ) \\
 & = & 4 \, \beta_{ghde} \, \beta_{cf} \, \left\{ 
	 {\cal J}^{cf} {\cal J}^{gd} {\cal J}^{he} 
	 + 2 \, {\cal J}^{cd} {\cal J}^{ge} {\cal J}^{hf} \right\} \,
	 g^{AB} \, \overline{\psi_A^a} \, ( D_{\mu} \psi_B^b ) \\
 & = & 4 \, \beta_{ghde} \, \beta_{cf} \, \left\{ 
	 g^{CF} g^{GD} g^{HE} + 2 \, g^{CD} g^{GE} g^{HF} \right\} \, 
         g^{AB} \\
 &   &	 \times \, \overline{\psi_C^c} \, {\psi_F^f} \, 
	 \overline{\psi_G^g} \, {\psi_D^d} \, \overline{\psi_H^h} \, 
	 {\psi_E^e} \, \overline{\psi_A^a} \, ( D_{\mu} \psi_B^b ) \\
 & = & 2 \, \beta_{ghde} \, \beta_{cf} \, \Big\{ 
	 g^{CF} g^{GD} g^{HE} + g^{CD} g^{GE} g^{HF} + g^{CE} g^{GF} 
         g^{HD} \\
& & - g^{CE} g^{GD} g^{HF} - g^{CF} g^{GE} g^{HD} - g^{CD} g^{GF} g^{HE}
	 \Big\} \, g^{AB} \\
 &   &	 \times \, \overline{\psi_C^c} \, {\psi_F^f} \, 
\overline{\psi_G^g} \,
	 {\psi_D^d} \, \overline{\psi_H^h} \, {\psi_E^e} \,
         \overline{\psi_A^a} \, ( D_{\mu} \psi_B^b ) \\
 & = & 2 \, \beta_{ghde} \, \beta_{cf} \, \overline{\epsilon}^{CGH} \,
	 \epsilon^{FDE} \, g^{AB} \,
	 \overline{\psi_C^c} \, {\psi_F^f} \, \overline{\psi_G^g} \,
	 {\psi_D^d} \, \overline{\psi_H^h} \, {\psi_E^e} \,
         \overline{\psi_A^a} \, ( D_{\mu} \psi_B^b ) \\
 & = & 2 \, \beta_{ghde} \, \beta_{cf} \, \overline{\epsilon}^{CGH} \,
	 \left\{ \epsilon^{FDB} \, g^{AE} + \epsilon^{DEB} \, g^{AF} +
	 \epsilon^{EFB} \, g^{AD} \right\} \\
 &   &	 \times \, \overline{\psi_C^c} \, {\psi_F^f} \, 
\overline{\psi_G^g} \,
	 {\psi_D^d} \, \overline{\psi_H^h} \, {\psi_E^e} \,
         \overline{\psi_A^a} \, ( D_{\mu} \psi_B^b ) \\
 & = & 2 \, \beta_{ghde} \, \beta_{cf} \, 
	 \left\{ 2 \, \overline{\epsilon}^{CGH} \, \epsilon^{FDB} \, g^{AE} 
	 + \overline{\epsilon}^{CGH} \, \epsilon^{DEB} \, g^{AF} \right\} \\
 &   &	 \times \, \overline{\psi_C^c} \, {\psi_F^f} \, 
\overline{\psi_G^g} \,
	 {\psi_D^d} \, \overline{\psi_H^h} \, {\psi_E^e} \,
         \overline{\psi_A^a} \, ( D_{\mu} \psi_B^b ) \\
 & = & 2 \, \beta_{ghde} \, \beta_{cf} \, 
	 \Big\{ 2 \, g^{CF} g^{GD} g^{HB} g^{AE} + 
2 \, g^{CD} g^{GB} g^{HF} g^{AE} 
	 + 2 \, g^{CB} g^{GF} g^{HD} g^{AE} \\
 &   &	 - 2 \, g^{CB} g^{GD} g^{HF} g^{AE} 
	 - 2 \, g^{CD} g^{GF} g^{HB} g^{AE} - 
2 \, g^{CF} g^{GB} g^{HD} g^{AE} \\
 &   &	 + g^{CD} g^{GE} g^{HB} g^{AF} + g^{CE} g^{GB} g^{HD} g^{AF} 
	 + g^{CB} g^{GD} g^{HE} g^{AF} \\
 &   &	 - g^{CB} g^{GE} g^{HD} g^{AF} 
	 - g^{CE} g^{GD} g^{HB} g^{AF} - g^{CD} g^{GB} g^{HE} g^{AF} 
           \Big\} \\
 &   &	 \times \, \overline{\psi_C^c} \, {\psi_F^f} \, 
\overline{\psi_G^g} \,
	 {\psi_D^d} \, \overline{\psi_H^h} \, {\psi_E^e} \,
         \overline{\psi_A^a} \, ( D_{\mu} \psi_B^b ) \\
 & = & - 4 \, \beta_{ghde} \, \beta_{cf} \, \big\{
	 2 \, {\cal J}^{cf} {\cal J}^{gd} {\cal J}^{ae} B_{\mu}^{hb}	   
       + 2 \, {\cal J}^{cd} {\cal J}^{hf} {\cal J}^{ae} B_{\mu}^{gb} \\
 &   & + 2 \, {\cal J}^{gf} {\cal J}^{hd} {\cal J}^{ae} B_{\mu}^{cb}	 
       + {\cal J}^{cd} {\cal J}^{ge} {\cal J}^{af} B_{\mu}^{hb} \\    
 &   & + {\cal J}^{ce} {\cal J}^{hd} {\cal J}^{af} B_{\mu}^{gb}       
       + {\cal J}^{gd} {\cal J}^{he} {\cal J}^{af} B_{\mu}^{cb} \big\} \\
 & = & - 4 \, B_{\mu}^{cb} \, \beta_{cf}
	 \left( {\cal J}^2 {\cal J}^{af} 
	 + 2 \, \beta_{ghde} \, {\cal J}^{gd} {\cal J}^{hf} {\cal J}^{ae} 
	 \right) \\
 &   & - 8 \, B_{\mu}^{gb} \, \beta_{ghde} \, \beta_{cf} \left(
	 {\cal J}^{he} {\cal J}^{cf} {\cal J}^{ad}     
	 + {\cal J}^{cd} {\cal J}^{hf} {\cal J}^{ae}
	 + {\cal J}^{hd} {\cal J}^{ce} {\cal J}^{af} \right).
\end{eqnarray*}
Finally, inserting (\ref{BtoC1}) we get (\ref{Id2}).
\hfill $\Box$ \vspace*{0.5cm}

Using the block representation of ${\cal J}^{ab}$ and $C_{\mu}^{ab}$
(see Appendix A), equation (\ref{Id2}) leads to four equations, written down 
in terms of spinor indices:
\begin{eqnarray}
{C_{\mu M}}^L \, Q({\cal J})^{\dot{K}M} & = & C_{\mu}^{\dot{M}L} \,
	\left( {\delta^{\dot{K}}}_{\dot{M}} \, {\cal X}^2 
	- {Q({\cal J})^{\dot{K}}}_{\dot{M}} \right)
      - (\partial_{\mu} {{\cal J}_M}^L) \, Q({\cal J})^{\dot{K}M} 
\nonumber \\
 &  & - (\partial_{\mu} {\cal J}^{\dot{M}L}) \, 
	{Q({\cal J})^{\dot{K}}}_{\dot{M}} 
	+ (\partial_{\mu} {\cal J}^{\dot{K}L}) \, {\cal X}^2 \label{Id2a} \\
{{C_{\mu}}^{\dot{M}}}_{\dot{L}} \, Q({\cal J})_{K \dot{M}} & = & 
	C_{\mu M \dot{L}} \, \left( {\delta_K}^M \, {\cal X}^2 
	- {Q({\cal J})_{K}}^M \right)
      - (\partial_{\mu} {\cal J}_{M \dot{L}}) \, {Q({\cal J})_K}^M 
\nonumber \\
 &  & - (\partial_{\mu} {{\cal J}^{\dot{M}}}_{\dot{L}}) \, 
	Q({\cal J})_{K \dot{M}} 
	+ (\partial_{\mu} {\cal J}_{K\dot{L}}) \, {\cal X}^2 \label{Id2b} \\
{C_{\mu}}^{\dot{M}L} \, Q({\cal J})_{K \dot{M}} & = & {C_{\mu M}}^L \,
	\left( {\delta_K}^M \, {\cal X}^2 
	- {Q({\cal J})_K}^M \right)
      - (\partial_{\mu} {{\cal J}_M}^L) \, {Q({\cal J})_K}^M \nonumber \\
 &  & - (\partial_{\mu} {\cal J}^{\dot{M}L}) \, 
	Q({\cal J})_{K \dot{M}} 
	+ (\partial_{\mu} {{\cal J}_K}^L) \, {\cal X}^2 \label{Id2c} \\   
C_{\mu M \dot{L}} \, Q({\cal J})^{\dot{K}M} & = & 
{{C_{\mu}}^{\dot{M}}}_{\dot{L}} \,
	\left( {\delta^{\dot{K}}}_{\dot{M}} \, {\cal X}^2 
	- {Q({\cal J})^{\dot{K}}}_{\dot{M}} \right)
      - (\partial_{\mu} {\cal J}_{M \dot{L}}) \, Q({\cal J})^{\dot{K}M} 
	\nonumber \\
 &  & - (\partial_{\mu} {{\cal J}^{\dot{M}}}_{\dot{L}}) \, 
	{Q({\cal J})^{\dot{K}}}_{\dot{M}}
	+ (\partial_{\mu} {{\cal J}^{\dot{K}}}_{\dot{L}}) \, 
{\cal X}^2 \label{Id2d} \\        
\end{eqnarray}
where
\begin{eqnarray}
{Q({\cal J})_K}^M & = & 8 \, \big( 
      - {\cal J}^2 {{\cal J}_K}^M
      + 2 \, {{\cal J}_O}^O {\cal J}^{\dot{N}M} {\cal J}_{K \dot{N}} 
      + 2 \, {{\cal J}^{\dot{N}}}_{\dot{N}} {{\cal J}_O}^M {{\cal J}_K}^O 
        \nonumber \\
 &  & + 2 \, {{\cal J}_O}^M {\cal J}^{\dot{N}O} {\cal J}_{K \dot{N}}
      + 2 \, {\cal J}_{O \dot{N}} {\cal J}^{\dot{N}M} {{\cal J}_K}^O
      + 2 \, {{\cal J}^{\dot{N}}}_{\dot{N}} {\cal J}^{\dot{O}M} 
{\cal J}_{K \dot{O}} \\
 &  & + {{\cal J}^{\dot{N}}}_{\dot{O}} {{\cal J}^{\dot{O}}}_{\dot{N}} 
{{\cal J}_K}^M
      + {{\cal J}^{\dot{O}}}_{\dot{O}} {{\cal J}^{\dot{N}}}_{\dot{N}} 
{{\cal J}_K}^M
      + 2 \, {\cal J}^{\dot{N}M} {{\cal J}^{\dot{O}}}_{\dot{N}} 
{\cal J}_{K \dot{O}}
      \big) \nonumber \\
{Q({\cal J})^{\dot{K}}}_{\dot{M}} & = & 8 \, \big( 
      - {\cal J}^2 {{\cal J}^{\dot{K}}}_{\dot{M}}
      + 2 \, {{\cal J}^{\dot{N}}}_{\dot{N}} {\cal J}_{O \dot{M}} 
{\cal J}^{\dot{K}O}
      + 2 \, {{\cal J}_O}^O {{\cal J}^{\dot{N}}}_{\dot{M}} 
{{\cal J}^{\dot{K}}}_{\dot{N}} 
        \nonumber \\
 &  & + 2 \, {{\cal J}^{\dot{N}}}_{\dot{M}} {\cal J}_{O \dot{N}} 
{\cal J}^{\dot{K}O}   
      + 2 \, {\cal J}_{O \dot{M}} {\cal J}^{\dot{N}O} 
{{\cal J}^{\dot{K}}}_{\dot{N}}
      + {{\cal J}_N}^N {{\cal J}_O}^O {{\cal J}^{\dot{K}}}_{\dot{M}} \\
 &  & + 2 \, {\cal J}_{N \dot{M}} {{\cal J}_O}^O {\cal J}^{\dot{K}N}
      + {{\cal J}_O}^N {{\cal J}_N}^O {{\cal J}^{\dot{K}}}_{\dot{M}}
      + 2 \, {\cal J}_{O \dot{M}} {{\cal J}_N}^O {\cal J}^{\dot{K}N}
      \big) \nonumber \\
Q({\cal J})^{\dot{K}M} & = & 8 \, \big( 
      - {\cal J}^2 {\cal J}^{\dot{K}M}
      + 2 \, {{\cal J}_O}^O {\cal J}^{\dot{N}M} 
{{\cal J}^{\dot{K}}}_{\dot{N}}
      + 2 \, {{\cal J}^{\dot{N}}}_{\dot{N}} {{\cal J}_O}^M 
{\cal J}^{\dot{K}O} 
        \nonumber \\
 &  & + 2 \, {{\cal J}_O}^M {\cal J}^{\dot{N}O} {{\cal J}^{\dot{K}}}_{\dot{N}}
      + 2 \, {\cal J}_{O \dot{N}} {\cal J}^{\dot{N}M} {\cal J}^{\dot{K}O}
      + 2 \, {{\cal J}^{\dot{N}}}_{\dot{N}} {\cal J}^{\dot{O}M} 
{{\cal J}^{\dot{K}}}_{\dot{O}} \\
 &  & + {{\cal J}^{\dot{N}}}_{\dot{O}} 
{{\cal J}^{\dot{O}}}_{\dot{N}} {\cal J}^{\dot{K}M}
      + {{\cal J}^{\dot{O}}}_{\dot{O}} 
{{\cal J}^{\dot{N}}}_{\dot{N}} {\cal J}^{\dot{K}M}
      + 2 \, {\cal J}^{\dot{N}M} 
{{\cal J}^{\dot{O}}}_{\dot{N}} {{\cal J}^{\dot{K}}}_{\dot{O}}
      \big) \nonumber \\
Q({\cal J})_{K \dot{M}} & = & 8 \, \big( 
      - {\cal J}^2 {\cal J}_{K \dot{M}}
      + 2 \, {{\cal J}^{\dot{N}}}_{\dot{N}} 
{\cal J}_{O \dot{M}} {{\cal J}_K}^O
      + 2 \, {{\cal J}_O}^O 
{{\cal J}^{\dot{N}}}_{\dot{M}} {\cal J}_{K \dot{N}} 
        \nonumber \\
 &  & + 2 \, {{\cal J}^{\dot{N}}}_{\dot{M}} 
{\cal J}_{O \dot{N}} {{\cal J}_K}^O 
      + 2 \, {\cal J}_{O \dot{M}} {\cal J}^{\dot{N}O} {\cal J}_{K \dot{N}}
      + {{\cal J}_N}^N {{\cal J}_O}^O {\cal J}_{K \dot{M}} \\
 &  & + 2 \, {\cal J}_{N \dot{M}} {{\cal J}_O}^O {{\cal J}_K}^N
      + {{\cal J}_O}^N {{\cal J}_N}^O {\cal J}_{K \dot{M}}
      + 2 \, {\cal J}_{O \dot{M}} {{\cal J}_N}^O {{\cal J}_K}^N
      \big). \nonumber 
\end{eqnarray}
Later on, two of these equations will be used to eliminate half of the
$C_{\mu}^{ab}$-fields under the functional integral. One can show by a 
straightforward calculation that all components of the $2 \times 2$-matrices 
${Q({\cal J})_K}^M$, 
${Q({\cal J})^{\dot{K}}}_{\dot{M}}$, $Q({\cal J})^{\dot{K}M}$ 
and $Q({\cal J})_{K \dot{M}}$ do not vanish identically.

Finally, we note that 
\begin{equation}
\label{Cmuab1}
C_{\mu}^{ab} = 2ig \, \overline{\psi_A^a} \, g^{AB} \psi_C^b \, {A_{\mu B}}^C
      + \left( \, \overline{\psi_A^a} \, g^{AB} \, (\partial_{\mu} \psi_B^b)
      + \psi_B^b \, g^{AB} \, (\partial_{\mu} \overline{\psi_A^a}) \, \right),
\end{equation}
which can be seen by inserting the covariant derivative (\ref{Dmu}) into the 
definition of $C_{\mu}^{ab}$.

\setcounter{equation}{0}
\section{The Lagrangian in Terms of Gauge Invariants}

To reformulate the Lagrangian (\ref{Lag}) in
terms of the gauge invariants introduced above, we use the 
same ideas as in the case of QED (see \cite{KRR}). In particular, for the
calculation of ${\cal L}_{gauge}$ we have to find a nonvanishing element
of maximal rank in the underlying Grassmann-algebra.

\begin{lemma}
\label{Lemma2}
The quantity $\left({\cal X}^2 \right)^4$  is a nonvanishing element 
of maximal rank in the Grass\-mann-algebra.
\end{lemma}
The proof of this Lemma is technical and can be found in Appendix B.

Next, let us introduce the following auxiliary variables convenient for
further calculations:
\begin{equation}
V^{Cab} := \epsilon^{ABC} \, \psi_A^a \, \psi_B^b. \label{V}
\end{equation}
Moreover, we denote 
$\psi^{Aa} \equiv g^{AB} \, \psi_B^a$.

\begin{lemma}
\label{Lemma1}
The following identity holds
\begin{equation}
\label{Id7}         
{\cal X}^2 \, g^{AB} = \overline{(\gamma_{\mu})_{gh}} \, 
	 \overline{\epsilon}^{\, CGH} \, \overline{\psi_H^h} \, 
	 \overline{\psi_G^g} \, \overline{\psi_C^c} \, \psi^{Ad} \, V^{Bef} \,
	 \{ 2 \, \beta_{ce} \, (\gamma^{\mu})_{fd} +
	 \beta_{cd} \, (\gamma^{\mu})_{fe} \}.
\end{equation}
\end{lemma}

{\sl Proof.}
To prove (\ref{Id7}), we make use of (\ref{eg}) as well as 
$V^{Cab} = V^{Cba}$ 
and the symmetry of $\beta_{ghde}$:
\begin{eqnarray*}
{\cal X}^2 \, g^{AB} 
 & = & 4 \, \beta_{ghde} \, \beta_{cf} \, \left\{ 
       {\cal J}^{cf} {\cal J}^{gd} {\cal J}^{he} 
       + 2 \, {\cal J}^{cd} {\cal J}^{ge} {\cal J}^{hf} \right\} \, g^{AB} \\
 & = & 4 \, \beta_{ghde} \, \beta_{cf} \, \left\{ 
       g^{CF} g^{GD} g^{HE} + 2 \, g^{CD} g^{GE} g^{HF} \right\} \, g^{AB} \,
       \overline{\psi_C^c} \, \psi_F^f \, \overline{\psi_G^g} \, \psi_D^d \, 
       \overline{\psi_H^h} \, \psi_E^e \\ 
 & = & 2 \, \beta_{ghde} \, \beta_{cf} \, \Big\{ 
       g^{CF} g^{GD} g^{HE} + g^{CD} g^{GE} g^{HF} + g^{CE} g^{GF} g^{HD} 
       - g^{CE} g^{GD} g^{HF} \\
 &   & - g^{CF} g^{GE} g^{HD} - g^{CD} g^{GF} g^{HE}
       \Big\} \, g^{AB} \, \overline{\psi_C^c} \, \psi_F^f \, 
       \overline{\psi_G^g} \, \psi_D^d \, \overline{\psi_H^h} \, \psi_E^e \\ 
 & = & 2 \, \beta_{ghde} \, \beta_{cf} \, \overline{\epsilon}^{\, CGH} \,
       \epsilon^{FDE} \, g^{AB} \, 
       \overline{\psi_C^c} \, \psi_F^f \, \overline{\psi_G^g} \, \psi_D^d \, 
       \overline{\psi_H^h} \, \psi_E^e \\ 
 & = & 2 \, \beta_{ghde} \, \beta_{cf} \, \overline{\epsilon}^{\, CGH} \,
       \left\{ \epsilon^{FDB} \, g^{AE} + \epsilon^{DEB} \, g^{AF} 
       + \epsilon^{EFB} \, g^{AD} \right\} \,  
       \overline{\psi_C^c} \, \psi_F^f \, \overline{\psi_G^g} \, \psi_D^d \, 
       \overline{\psi_H^h} \, \psi_E^e \\ 
 & = & \overline{(\gamma_{\mu})_{gh}} \, (\gamma^{\mu})_{de} \, \beta_{cf} \,
       \overline{\epsilon}^{\, CGH} \, \left\{
       2 \, \epsilon^{FDB} \, g^{AE} + \epsilon^{DEB} \, g^{AF} \right\} \,
       \overline{\psi_C^c} \, \psi_F^f \, \overline{\psi_G^g} \, \psi_D^d \, 
       \overline{\psi_H^h} \, \psi_E^e \\ 
 & = & \overline{(\gamma_{\mu})_{gh}} \, \overline{\epsilon}^{\, CGH} \, 
       \overline{\psi_H^h} \, \overline{\psi_G^g} \, \overline{\psi_C^c} \,
       \beta_{cf} \, (\gamma^{\mu})_{de} \, \left\{
       2 \, V^{Bfd} \, \psi_E^e \, g^{AE} + V^{Bde} \, \psi_F^f \, g^{AF} 
       \right\} \\
 & = & \overline{(\gamma_{\mu})_{gh}} \, \overline{\epsilon}^{\, CGH} \, 
       \overline{\psi_H^h} \, \overline{\psi_G^g} \, \overline{\psi_C^c} \,
       \psi^{Ad} \, V^{Bef} \, \left\{
       2 \, \beta_{ce} \, (\gamma^{\mu})_{fd} + 
\beta_{cd} \, (\gamma^{\mu})_{fe} 
       \right\}.
\end{eqnarray*}       
\hfill $\Box$ \vspace*{0.5cm}

\begin{proposition}
\label{Prop1}
We have
\begin{equation}
\label{LagId1}
({\cal X}^2)^4 \, {\cal L}_{gauge} 
 = \frac{1}{8} \left( G_{\mu\nu} \right)_{ab} \,
   \left( G^{\mu\nu} \right)_{cd} \, \epsilon^{bc} \, \epsilon^{da} 
\end{equation}
with
\begin{eqnarray}
\label{LagId2}
\left( G_{\mu\nu} \right)_{ab}	
& := & \frac{2 \, {\cal X}^2}{i g} \, \epsilon_{ha} \, \beta_{cdefgb} \, 
       {\cal J}^{cf} {\cal J}^{dg} ( \partial_{[\mu} C_{\nu]}^{eh} ) 
\nonumber\\
&   &  + \frac{4}{i g} \, \epsilon_{ha} \, \beta_{cdefgb} \, 
\beta_{klmnop} \, 
       {\cal J}^{cf} {\cal J}^{dg} {\cal J}^{kn} {\cal J}^{lo} \\
&   &  \hspace*{0.7cm} \times \, \big\{ (\partial_{[\mu} 
{\cal J}^{ep}) (\partial_{\nu]} {\cal J}^{mh})
       - C_{[\mu}^{ep} \, (\partial_{\nu]} {\cal J}^{mh}) 
       + (\partial_{[\mu} {\cal J}^{ep}) \, C_{\nu]}^{mh}
       - C_{[\mu}^{ep} \, C_{\nu]}^{mh} \big\}, \nonumber
\end{eqnarray}
where
\begin{equation}
\label{betanew}
\beta_{abcdef} := 2 \, \{ \beta_{abde} \beta_{cf} 
		+ 2 \, \beta_{abef} \beta_{cd} 
		+ \beta_{bcde} \beta_{af} + 2 \, \beta_{bcef} \beta_{ad} 
		+ \beta_{cade} \beta_{bf} + 2 \, \beta_{caef} \beta_{bd} \}.
\end{equation}
Moreover, the matter Lagrangian ${\cal L}_{mat}$ takes the form
\begin{equation}
\label{LagId3}
{\cal L}_{mat} = - m \, \beta_{ab} \, {\cal J}^{ab} 
    - \frac{1}{2} \, {\rm Im} \left\{ 
      \beta_{ab} \, {(\gamma^{\mu})^b}_c \, (\partial_{\mu} {\cal J}^{ac})
      + \beta_{ab} \, {(\gamma^{\mu})^b}_c \, C_{\mu}^{ac} \right\}. 
\end{equation}
\end{proposition}        

{\sl Proof.}
To prove this Proposition we make use of identity (\ref{Id7}):
\begin{eqnarray*}
\lefteqn{({\cal X}^2)^4 \, {\cal L}_{gauge}} \\
 & = & - \frac{1}{8} \, ({\cal X}^2)^4 \, 
       {{F_{\mu \nu}}_A}^B {{F^{\mu \nu}}_B}^A, \\
 & = & - \frac{1}{8} \, ({\cal X}^2)^2 \,
       \left( {{F_{\mu \nu}}_A}^B {\cal X}^2 \right)
       \left( {{F^{\mu \nu}}_B}^A {\cal X}^2 \right) \\
 & = & - \frac{1}{8} \, ({\cal X}^2)^2 \,	
       \left( F_{\mu\nu AC} \, g^{CB} \, {\cal X}^2 \right)
       \left( {F^{\mu\nu}}_{BD} \, g^{DA} \, {\cal X}^2 \right) \\
 & = & - \frac{1}{8} \, ({\cal X}^2)^2 \, \left( F_{\mu\nu AC} \, 
       \overline{(\gamma_{\sigma})_{fg}} \, \overline{\epsilon}^{EFG} \,
       \overline{\psi_G^g} \, \overline{\psi_F^f} \, \overline{\psi_E^e} \,
       \psi^C_m \, \epsilon^{cm} \, V^{Bab} \, \{ 2 \, \beta_{ea} \, 
(\gamma^{\sigma})_{bc} + \beta_{ec} \, (\gamma^{\sigma})_{ba} \} \right) \\
 &   & \times \, \left( F^{\mu\nu}_{BD} \, 
       \overline{(\gamma_{\delta})_{ij}} \, \overline{\epsilon}^{HIJ} \,
       \overline{\psi_J^j} \, \overline{\psi_I^i} \, \overline{\psi_H^h} \,
       \psi^D_n \, \epsilon^{dn} \, V^{Akl} \, \{ 2 \,
 \beta_{hk} \, (\gamma^{\delta})_{ld} + 
       \beta_{hd} \, (\gamma^{\delta})_{lk} \} \right).
\end{eqnarray*}
A reordering of factors yields
\begin{eqnarray}
\label{res1}
({\cal X}^2)^4 \, {\cal L}_{gauge} & = &
       \frac{1}{8} \, \left( {\cal X}^2 \, V^{Akl} \,
       \overline{(\gamma_{\delta})_{ij}} \, \overline{\epsilon}^{HIJ} \,
       \overline{\psi_J^j} \, \overline{\psi_I^i} \, \overline{\psi_H^h} \,
       F_{\mu\nu AC} \, \psi^C_m \, \gamma^{\delta}_{hkld} \right) \, 
       \epsilon^{cm} \nonumber \\
 &  &  \times \, \left( {\cal X}^2 \, V^{Bab} \, 
       \overline{(\gamma_{\sigma})_{fg}} \, \overline{\epsilon}^{EFG} \,
       \overline{\psi_G^g} \, \overline{\psi_F^f} \, \overline{\psi_E^e} \,
       {F^{\mu\nu}}_{BD} \, \psi^D_n \, \gamma^{\sigma}_{eabc} \right) \,
       \epsilon^{dn} \nonumber \\
 & =:& \frac{1}{8} \, \left( G_{\mu\nu} \right)_{md} \, \epsilon^{cm} \,
       \left( G^{\mu\nu} \right)_{nc} \, \epsilon^{dn}, 
\end{eqnarray}
where we introduced
$$
\label{gammanew}
\gamma^{\mu}_{abcd} := 2 \, \beta_{ab} \, (\gamma^{\mu})_{cd}
       + \beta_{ad} \, (\gamma^{\mu})_{cb}.
$$
It remains to calculate $(G_{\mu\nu})_{md}$. Using
\begin{eqnarray*}
F_{\mu\nu AC} \psi^{Cb} 
 & \equiv & \frac{1}{ig} \, (D_{[\mu} D_{\nu]} \psi_A^b),
\end{eqnarray*}
we obtain
\begin{eqnarray*}
\left( G_{\mu\nu} \right)_{md}
 & = & \lefteqn{ {\cal X}^2 \, V^{Akl} \, 
       \overline{(\gamma_{\delta})_{ij}} \, \overline{\epsilon}^{HIJ} \,
       \overline{\psi_J^j} \, \overline{\psi_I^i} \, \overline{\psi_H^h} \,
       F_{\mu\nu AC} \, \psi^C_m \, \gamma^{\delta}_{hkld}  } \\
 & = & \frac{1}{ig} \, {\cal X}^2 \, V^{Akl} \, 
       \overline{(\gamma_{\delta})_{ij}} \, \overline{\epsilon}^{HIJ} \,
       \overline{\psi_J^j} \, \overline{\psi_I^i} \, \overline{\psi_H^h} \,
       \, (D_{[\mu} D_{\nu]} \psi_A^a) \, (\gamma^{\delta}_{hkld})
       \, \epsilon_{am} \\
 & = & \frac{2}{ig} \, {\cal X}^2 \, \gamma^{\delta}_{hkld} \, 
\epsilon_{am} \,
       \overline{(\gamma_{\delta})_{ij}} \,
       \psi^k_K \, \psi^l_L \, \overline{\psi^j_J} \,
       \overline{\psi^i_I} \, \overline{\psi^h_H} \, 
       \left( D_{[\mu} D_{\nu]} \psi^a_A \right) \\
 &   & \times \, \{ g^{HK} g^{IL} g^{JA} + g^{HL} g^{IA} g^{JK} 
       + g^{HA} g^{IK} g^{JL} \}.
\end{eqnarray*}
In the last step the definition of $V^{Cab}$ and the identity
(\ref{ee}) were inserted. Moreover, the antisymmetry in the color
indices $(I,J)$ and the symmetry in the bispinor indices $(i,j)$ were
used. Changing the indices in each term of the above sum separately,
the last equation gives
\begin{eqnarray*}
\left( G_{\mu\nu} \right)_{md} 
 & = & \frac{2}{ig} \, {\cal X}^2 \, g^{HK} g^{IL} g^{JA} \,
       \psi^k_K \, \psi^l_L \, \overline{\psi^j_J} \, \overline{\psi^i_I} \,
       \overline{\psi^h_H} \, \left( D_{[\mu} D_{\nu]} \psi^a_A \right) \\
 &   & \times \, \{ \gamma^{\delta}_{hkld} \, 
\overline{(\gamma_{\delta})_{ij}} 
       + \gamma^{\delta}_{ikld} \, \overline{(\gamma_{\delta})_{jh}}
       + \gamma^{\delta}_{jkld} \, \overline{(\gamma_{\delta})_{hi}} \, \} 
       \, \epsilon_{am} \\ 
 & = & \frac{2}{ig} \, {\cal X}^2 \, {\cal J}^{hk} {\cal J}^{il} \,
       \overline{\psi^j_J} \, g^{JA} 
\left( D_{[\mu} D_{\nu]} \psi^a_A \right)	 
       \, \beta_{hijkld} \, \epsilon_{am},
\end{eqnarray*}
where $\beta_{hijkld}$ is given by equation (\ref{betanew}). With
(\ref{B}) and (\ref{Id7}) we get 
\begin{eqnarray*}
\left( G_{\mu\nu} \right)_{md} 
 & = & \frac{2}{ig} \, {\cal X}^2 \, {\cal J}^{hk} {\cal J}^{il} 
       \left\{ D_{[\mu} \left( \overline{\psi^j_J} \, g^{JA} (D_{\nu]} 
\psi^a_A) \right)
       - (D_{[\mu} \overline{\psi^j_J}) \, g^{JA} (D_{\nu]} \psi^a_A) 
\right\} \,
       \beta_{hijkld} \, \epsilon_{am} \\
 & = & \frac{2}{ig} \, {\cal X}^2 \, {\cal J}^{hk} {\cal J}^{il} 
       (D_{[\mu} B_{\nu]}^{ia}) \, \beta_{hijkld} \, \epsilon_{am} \\
 &   & - \frac{2}{ig} \, {\cal X}^2 \, g^{JA} \, {\cal J}^{hk} {\cal J}^{il} 
       (D_{[\mu} \overline{\psi^j_J}) \, (D_{\nu]} \psi^a_A) \,
       \beta_{hijkld} \, \epsilon_{am} \\
 & = & \frac{2}{ig} \, {\cal X}^2 \, {\cal J}^{hk} {\cal J}^{il} 
       (D_{[\mu} B_{\nu]}^{ia}) \, \beta_{hijkld} \, \epsilon_{am} \\
 &   & - \frac{2}{ig} \, \overline{(\gamma_{\sigma})_{fg}} \, 
       \overline{\epsilon}^{\, EFG} \, \overline{\psi_G^g} \, 
\overline{\psi_F^f} 
       \, \overline{\psi_E^e} \, \psi^{Jn} \, V^{Aop} \,
       \{ 2 \, \beta_{eo} \, (\gamma^{\mu})_{pn} +
       \beta_{en} \, (\gamma^{\sigma})_{po} \} \\ 
 &   & \hspace*{0.75cm} \times \, {\cal J}^{hk} {\cal J}^{il} 
       (D_{[\mu} \overline{\psi^j_J}) \, (D_{\nu]} \psi^a_A) \,
       \beta_{hijkld} \, \epsilon_{am} \\
 & = & \frac{2}{ig} \, {\cal X}^2 \, {\cal J}^{hk} {\cal J}^{il} 
       (D_{[\mu} B_{\nu]}^{ia}) \, \beta_{hijkld} \, \epsilon_{am} \\
 &   & + \frac{2}{ig} \, {\cal J}^{hk} {\cal J}^{il} \, 
       \overline{\epsilon}^{EFG} \, \epsilon^{BCA} \,
       \overline{\psi_G^g} \, \overline{\psi_F^f} \, \overline{\psi_E^e} \, 
       \psi_B^b \, \psi_C^c \, \overline{B_{[\mu}^{nj}} \, (D_{\nu]} \psi^a_A) 
       \, \overline{(\gamma_{\sigma})_{fg}} \, 
       \beta_{hijkld} \, \epsilon_{am} \, \gamma^{\sigma}_{eopn} \\
 & = & \frac{2}{ig} \, {\cal X}^2 \, {\cal J}^{hk} {\cal J}^{il} 
       (D_{[\mu} B_{\nu]}^{ia}) \, \beta_{hijkld} \, \epsilon_{am} \\
 &   & + \frac{4}{ig} \, {\cal J}^{hk} {\cal J}^{il} \, 
g^{EB} g^{FC} g^{GA} \,
       \overline{\psi_G^g} \, \overline{\psi_F^f} \, 
\overline{\psi_E^e} \,				
       \psi_B^b \, \psi_C^c \, \overline{B_{[\mu}^{nj}} \, 
(D_{\nu]} \psi^a_A) \,
       \beta_{hijkld} \, \beta_{efgbcn} \, \epsilon_{am},
\end{eqnarray*}
and, finally,
\begin{eqnarray}
\label{res2}
\left( G_{\mu\nu} \right)_{md}
 & = & \frac{2}{ig} \, {\cal X}^2 \, {\cal J}^{hk} {\cal J}^{il} 
       (\partial_{[\mu} B_{\nu]}^{ia}) \, \beta_{hijkld} \, 
\epsilon_{am} \nonumber \\
 &   & + \frac{4}{ig} \, {\cal J}^{hk} {\cal J}^{il} 
       {\cal J}^{eb} {\cal J}^{fc} \, \overline{B_{[\mu}^{nj}} \, 
B_{\nu]}^{ga} 
       \, \beta_{hijkld} \, \beta_{efgbcn} \, \epsilon_{am}.
\end{eqnarray}
The last step consists in the elimination of the auxiliary variables
$B_{\mu}^{ab}$.  Inserting (\ref{BtoC1}) and (\ref{BtoC2}) into
(\ref{res2}) and renaming some indices we get (\ref{LagId2}). Together
with (\ref{res1}) this completes the proof of (\ref{LagId1}).

To show (\ref{LagId3}) we simply insert the definition of ${\cal
J}^{ab}$ and $B_{\mu}^{ab}$ into ${\cal L}_{mat}$:
\begin{eqnarray*}
{\cal L}_{mat} 
 & = & - m \, \overline{\psi^a_A} \, \beta_{ab} \, g^{AB} \, \psi^b_B 
       - {\rm Im} \left\{ \overline{\psi^a_A} \, \beta_{ab} \,
	 {(\gamma^{\mu})^b}_c \, g^{AB} \, 
         (D_{\mu}\psi^c_B) \right\} \\
 & = & - m \, \beta_{ab} \, {\cal J}^{ab}
       - {\rm Im} \{ \beta_{ab} \, {(\gamma^{\mu})^b}_c \, B_{\mu}^{ac} \}
\end{eqnarray*}
Using again (\ref{BtoC1}) to eliminate $B_{\mu}^{ab}$, we obtain equation 
(\ref{LagId3}).
\hfill $\Box$ \vspace*{0.5cm} \\
{\bf Remarks:}
\begin{enumerate}
\item
Formula (\ref{LagId1}) is an identity on the level of elements of
maximal rank in the underlying Grassmann-algebra. Since the space of
elements of maximal rank is one-dimensional, dividing by a
non-vanishing element of maximal rank is a well-defined operation
giving a c-number. Thus, knowing $({\cal X}^2)^4 \, {\cal L}_{gauge}$,
we can reconstruct ${\cal L}_{gauge}$ uniquely just dividing by $({\cal
X}^2)^4$. This means that Proposition 2 gives us the Lagrangian in
terms of invariants ${\cal J}^{ab}$ and $C_{\mu}^{ab}$.

\item
The additional algebraic identities (\ref{Id2}), which are basic for
getting the correct number of degrees of freedom, cannot be ``solved''
on the level of the algebra of Grassmann-algebra-valued invariants.
However, as will be shown in the next section, it is possible to
implement them under the functional integral. This enables us to
eliminate half the number of components of $C_{\mu}^{ab}$. The result
will be an effective functional integral in terms of the correct number
of degrees of freedom, see also the Introduction for a discussion of
this point.
\end{enumerate}

\setcounter{equation}{0}
\section{The Functional Integral}

Now we start to reformulate the functional integral (\ref{FunInt1}). For that
purpose we will use the following notion of the $\delta$-distribution on
superspace
\begin{equation}
\label{delta1}
\delta(u-U) = \int e^{2 \pi i \, \xi (u-U)} \, {\rm d} \xi 
	    = \sum_{n=0}^{\infty} \, \frac{(-1)^n}{n!} \, 
	      \delta^{(n)}(u) U^n, 
\end{equation}
where $u$ is a c-number variable and $U$ a combination of Grassmann
variables $\psi$ and $\overline{\psi}$ with rank smaller or equal to
the maximal rank. Due to the nilpotent character of $U$ the above sum
is finite. This $\delta$-distribution is a special example of a
vector-space-valued distribution in the sense of
\cite{T}. From the above definition we have immediately
\begin{equation}
\label{delta2}
1 = \int \delta(u-U) \, {\rm d}u.
\end{equation}

One easily shows the following
\begin{lemma}
For an arbitrary smooth function $f$ we have
\begin{equation}
\label{delta3}
f(u) \, \delta(u-U) = f(U) \, \delta(u-U).
\end{equation}
\end{lemma}
The above observation leads to a technique, which frequently will be
used in this section:
\begin{eqnarray}
\label{delta4}
\int f(\alpha) \, \delta(u-U) \, {\rm d}u 
 & = & \int f \left( \alpha \, \frac{g(u)}{g(\tilde{u})} \right) \, 
\delta(u-U) \, 
       \delta(u-\tilde{u}) \, {\rm d}u \, {\rm d}{\tilde{u}} \nonumber \\
 & = & \int f \left( \frac{\alpha \, g(U)}{g(u)} \right) \, \delta(u-U) \, 
       \delta(U-\tilde{u}) \, {\rm d}u \, {\rm d}{\tilde{u}}.
\end{eqnarray}
Here, $\alpha$ denotes any c-number or Grassmann-algebra valued quantity and 
$g(u)$ is an arbitrary smooth function of $u$, such that the rank of 
$\alpha \, g(U)$ is smaller or equal to the maximal rank of the underlying 
Grassmann-algebra.

Now we introduce new, independent fields $(j,c)$ associated with the
invariants $({\cal J}^{ab}$, $C_{\mu}^{ab})$ in a sense, which will
become clear in what follows. Both fields $j$ and $c$ are by definition
bosonic and gauge-invariant; $j=(j^{ab})$ is a (c-number-valued)
Hermitean spin tensor field of second rank and $c=(c_{\mu}^{ab})$ is a
(c-number-valued) anti-Hermitean spin-tensor-valued covector field. We
call $(j,c)$ c-number mates of $(J,C)$.

As mentioned earlier, equations (\ref{Id2a}) -- (\ref{Id2d}) can be
used to substitute half the number of components of the covector field
$C_{\mu}^{ab}$.  We choose ${C_{\mu K}}^L$ as independent variables.
Thus (\ref{Id2c}) and (\ref{Id2d}) can be used to eliminate all
components in the diagonal blocks of $C_{\mu}^{ab}$ (see Appendix A).
In the subspace of our bispinorspace, which corresponds to the elements
of the off-diagonal blocks of the field $C_{\mu}^{ab}$, we choose the
following basis elements $e_{\rho}$, $\rho = 1 \ldots 8$:
\begin{eqnarray}
& e_1 = \left( \begin{array}{cccc} 
    0 & 0 & 1 & 0 \\
    0 & 0 & 0 & 0 \\
    0 & 0 & 0 & 0 \\
    0 & 0 & 0 & 0 \end{array} \right)
\quad 
e_2 = \left( \begin{array}{cccc} 
    0 & 0 & 0 & 1 \\
    0 & 0 & 0 & 0 \\
    0 & 0 & 0 & 0 \\
    0 & 0 & 0 & 0 \end{array} \right) 
\quad
e_3 = \left( \begin{array}{cccc} 
    0 & 0 & 0 & 0 \\
    0 & 0 & 1 & 0 \\
    0 & 0 & 0 & 0 \\
    0 & 0 & 0 & 0 \end{array} \right) & \nonumber \\
& e_4 = \left( \begin{array}{cccc} 
    0 & 0 & 0 & 0 \\
    0 & 0 & 0 & 1 \\
    0 & 0 & 0 & 0 \\
    0 & 0 & 0 & 0 \end{array} \right)
\quad 
e_5 = \left( \begin{array}{cccc} 
    0 & 0 & 0 & 0 \\
    0 & 0 & 0 & 0 \\
    1 & 0 & 0 & 0 \\
    0 & 0 & 0 & 0 \end{array} \right) 
\quad
e_6 = \left( \begin{array}{cccc} 
    0 & 0 & 0 & 0 \\
    0 & 0 & 0 & 0 \\
    0 & 1 & 0 & 0 \\
    0 & 0 & 0 & 0 \end{array} \right) & \label{spinorbasis} \\
& e_7 = \left( \begin{array}{cccc} 
    0 & 0 & 0 & 0 \\
    0 & 0 & 0 & 0 \\
    0 & 0 & 0 & 0 \\
    1 & 0 & 0 & 0 \end{array} \right)
\quad 
e_8 = \left( \begin{array}{cccc} 
    0 & 0 & 0 & 0 \\
    0 & 0 & 0 & 0 \\
    0 & 0 & 0 & 0 \\
    0 & 1 & 0 & 0 \end{array} \right). & \nonumber
\end{eqnarray}
A basis in colorspace is given by the Gell--Mann--matrices $t_{\alpha}$, 
where $\alpha = 1 \ldots 8$:
\begin{eqnarray}
& t_1 = \left( \begin{array}{ccc} 
    0 & 1 & 0 \\
    1 & 0 & 0 \\
    0 & 0 & 0 \end{array} \right) 
\quad 
t_2 = \left( \begin{array}{ccc} 
    0 & -i & 0 \\
    i &  0 & 0 \\
    0 &  0 & 0 \end{array} \right) 
\quad	 
t_3 = \left( \begin{array}{ccc} 
    1 &  0 & 0 \\
    0 & -1 & 0 \\
    0 &  0 & 0 \end{array} \right) & \nonumber \\
& t_4 = \left( \begin{array}{ccc} 
    0 & 0 & 1 \\
    0 & 0 & 0 \\
    1 & 0 & 0 \end{array} \right) 
\quad	  
t_5 = \left( \begin{array}{ccc} 
    0 & 0 & -i \\
    0 & 0 &  0 \\
    i & 0 &  0 \end{array} \right) 
\quad	 
t_6 = \left( \begin{array}{ccc} 
    0 & 0 & 0 \\
    0 & 0 & 1 \\
    0 & 1 & 0 \end{array} \right) & \label{colorbasis} \\
& t_7 = \left( \begin{array}{ccc} 
    0 & 0 &  0 \\
    0 & 0 & -i \\
    0 & i &  0 \end{array} \right) 
\quad	  
t_8 = \frac{1}{\sqrt{3}} \, \left( \begin{array}{ccc} 
    1 & 0 &  0 \\
    0 & 1 &  0 \\
    0 & 0 & -2 \end{array} \right). & \nonumber
\end{eqnarray}
Moreover, we denote
\begin{eqnarray}
& \chi^2 = 4 \, j^{ad} j^{be} j^{cf} \,
       \left\{ \beta_{bcef} \, \beta_{ad} + 2 \, \beta_{bcde} \, 
       \beta_{af} \right\}, & \label{chi}, \\				   
& j^2 = j^{ab} \, \beta_{acbd} \, j^{cd}, & \label{j}	   
\end{eqnarray}

\begin{equation}
{M(j)_{\rho}}^{\sigma} = 4 g^2 \, \left\{ \frac{2}{3} \, {j^a}_b \, {j^c}_d
	   + 2 \, {j^a}_d \, {j^c}_b \right\} \, {(e_{\rho})_a}^b \, 
	   {(e^{\sigma})_c}^d,
\end{equation}
\begin{eqnarray}            
{Q(j)_K}^M & = & 8 \, \Big( 
      - j^2 {j_K}^M + 2 \, {j_O}^O j^{\dot{N}M} j_{K \dot{N}} 
      + 2 \, {j^{\dot{N}}}_{\dot{N}} {j_O}^M {j_K}^O \nonumber \\
 &  & + 2 \, {j_O}^M j^{\dot{N}O} j_{K \dot{N}} 
      + 2 \, j_{O \dot{N}} j^{\dot{N}M} {j_K}^O
      + 2 \, {j^{\dot{N}}}_{\dot{N}} j^{\dot{O}M} j_{K \dot{O}} \\
 &  & + {j^{\dot{N}}}_{\dot{O}} {j^{\dot{O}}}_{\dot{N}} {j_K}^M 
      + {j^{\dot{O}}}_{\dot{O}} {j^{\dot{N}}}_{\dot{N}} {j_K}^M
      + 2 \, j^{\dot{N}M} {j^{\dot{O}}}_{\dot{N}} j_{K \dot{O}}
      \Big), \nonumber \\
{Q(j)^{\dot{K}}}_{\dot{M}} & = & 8 \, \Big( 
      - j^2 {j^{\dot{K}}}_{\dot{M}}
      + 2 \, {j^{\dot{N}}}_{\dot{N}} j_{O \dot{M}} j^{\dot{K}O}
      + 2 \, {j_O}^O {j^{\dot{N}}}_{\dot{M}} {j^{\dot{K}}}_{\dot{N}} 
\nonumber \\
 &  & + 2 \, {j^{\dot{N}}}_{\dot{M}} j_{O \dot{N}} j^{\dot{K}O}
      + 2 \, j_{O \dot{M}} j^{\dot{N}O} {j^{\dot{K}}}_{\dot{N}}
      + {j_N}^N {j_O}^O {j^{\dot{K}}}_{\dot{M}} \\
 &  & + 2 \, j_{N \dot{M}} {j_O}^O j^{\dot{K}N}
      + {j_O}^N {j_N}^O {j^{\dot{K}}}_{\dot{M}}
      + 2 \, j_{O \dot{M}} {j_N}^O j^{\dot{K}N}
      \Big), \nonumber \\
Q(j)^{\dot{K}M} & = & 8 \, \Big( 
      - j^2 j^{\dot{K}M}
      + 2 \, {j_O}^O j^{\dot{N}M} {j^{\dot{K}}}_{\dot{N}}
      + 2 \, {j^{\dot{N}}}_{\dot{N}} {j_O}^M j^{\dot{K}O} \nonumber \\
 &  & + 2 \, {j_O}^M j^{\dot{N}O} {j^{\dot{K}}}_{\dot{N}}
      + 2 \, j_{O \dot{N}} j^{\dot{N}M} j^{\dot{K}O}
      + 2 \, {j^{\dot{N}}}_{\dot{N}} j^{\dot{O}M} {j^{\dot{K}}}_{\dot{O}} \\
 &  & + {j^{\dot{N}}}_{\dot{O}} {j^{\dot{O}}}_{\dot{N}} j^{\dot{K}M}
      + {j^{\dot{O}}}_{\dot{O}} {j^{\dot{N}}}_{\dot{N}} j^{\dot{K}M}
      + 2 \, j^{\dot{N}M} {j^{\dot{O}}}_{\dot{N}} {j^{\dot{K}}}_{\dot{O}}
      \Big), \nonumber \\
Q(j)_{K \dot{M}} & = & 8 \, \Big( 
      - j^2 j_{K \dot{M}}
      + 2 \, {j^{\dot{N}}}_{\dot{N}} j_{O \dot{M}} {j_K}^O
      + 2 \, {j_O}^O {j^{\dot{N}}}_{\dot{M}} j_{K \dot{N}} \nonumber \\
 &  & + 2 \, {j^{\dot{N}}}_{\dot{M}} j_{O \dot{N}} {j_K}^O 
      + 2 \, j_{O \dot{M}} j^{\dot{N}O} j_{K \dot{N}}
      + {j_N}^N {j_O}^O j_{K \dot{M}} \\
 &  & + 2 \, j_{N \dot{M}} {j_O}^O {j_K}^N
      + {j_O}^N {j_N}^O j_{K \dot{M}}
      + 2 \, j_{O \dot{M}} {j_N}^O {j_K}^N
      \Big). \nonumber 
\end{eqnarray}

\begin{proposition}
The functional integral ${\cal F}$ in terms of the gauge invariant set 
$(j, c)$ is given by
\begin{equation}
\label{GIFunInt}
{\cal F} = \int \prod {\rm d}j \, {\rm d}c \, K[j] \, {\rm e}^{i S[j,c]},
\end{equation}
with the integral kernel
\begin{eqnarray}
K[j] & = & \frac{6^4}{12!} \,
	 \frac{(\chi^2)^4}{{\rm det}[{M(j)_{\rho}}^{\sigma}] \, (j^2)^2} \,
	 \epsilon_{c_1 c_4 c_7 c_{10}} \, \epsilon_{c_2 c_5 c_8 c_{11}} \,
	 \epsilon_{c_3 c_6 c_9 c_{12}} \, \epsilon^{b_1 b_4 b_7 b_{10}} \,
	 \epsilon^{b_2 b_5 b_8 b_{11}} \, \epsilon^{b_3 b_6 b_9 b_{12}} 
\nonumber \\
  &   &  \times \, \left\{ \prod_{r=1}^{12} \beta^{a_r c_r} \, 
	 \frac{\partial}{\partial j^{a_r b_r}} \right\} \, \delta(j) 
\end{eqnarray}
and
\begin{equation}
\label{new}
S[j,c] = \int {\rm d}^4 x \,  {\cal L} [j,c] \ ,
\end{equation}
with the effective Lagrangian ${\cal L} [j,c]$ given by
\begin{eqnarray}
\label{GILag}
{\cal L}[j,c]
  & = &  \frac{1}{8 \, (\chi^2)^4} \left( G_{\mu\nu} \right)_{ab} \,
	 \left( G^{\mu\nu} \right)_{cd} \, \epsilon^{bc} \, \epsilon^{da} 
	 - m \, \beta_{ab} \, j^{ab} \nonumber \\
  &   &  - \frac{1}{2} \, {\rm Im} \left\{ 
	 \beta_{ab} \, {(\gamma^{\mu})^b}_c \, (\partial_{\mu} j^{ac})
	 + \beta_{ab} \, {(\gamma^{\mu})^b}_c \, c_{\mu}^{ac} \right\} \, ,
\end{eqnarray}
where
\begin{eqnarray}
\left( G_{\mu\nu} \right)_{ab}	
 & = & \frac{2 \, \chi^2}{i g} \, \epsilon_{ha} \, \beta_{cdefgb} \, 
       j^{cf} j^{dg} ( \partial_{[\mu} c_{\nu]}^{eh} ) \nonumber\\
& & + \frac{4}{i g} \, \epsilon_{ha} \, \beta_{cdefgb} \,
\beta_{klmnop} \, 
       j^{cf} j^{dg} j^{kn} j^{lo} \\
 &   & \hspace*{0.7cm} \times \, 
       \big\{ (\partial_{[\mu} j^{ep}) (\partial_{\nu]} j^{mh})
       - c_{[\mu}^{ep} \, (\partial_{\nu]} j^{mh}) 
       + (\partial_{[\mu} j^{ep}) \, c_{\nu]}^{mh}
       - c_{[\mu}^{ep} \, c_{\nu]}^{mh} \big\} \, . \nonumber
\end{eqnarray}
Moreover, among the quantities $c_{\mu}^{ab}$ only the ${c_{\mu K}}^L$ 
(and their complex conjugate 
${{c_{\mu}}^{\dot{L}}}_{\dot{K}} = \overline{{c_{\mu K}}^L}$)
are independent. The remaining quantities, $c_{\mu K \dot{L}}$ and
${c_{\mu}}^{\dot{K}L}$,
have to be eliminated in (\ref{GILag}) due to the following identities:
\begin{eqnarray}
c_{\mu K \dot{L}} & = & (Q(j)^{-1})_{K \dot{M}} \Big\{
	    \left( {\delta^{\dot{M}}}_{\dot{N}} \, 
{\chi}^2 - {Q(j)^{\dot{M}}}_{\dot{N}} \right) 
	    {{c_{\mu}}^{\dot{N}}}_{\dot{L}} 
	  - (\partial_{\mu} j_{N \dot{L}}) \, Q(j)^{\dot{M}N} \nonumber \\
   &  &   - (\partial_{\mu} {j^{\dot{N}}}_{\dot{L}}) \, 
	    {Q(j)^{\dot{M}}}_{\dot{N}} 
	    + ({\partial_{\mu}} {j^{\dot{M}}}_{\dot{L}}) \, {\chi}^2 \Big\} \\
{c_{\mu}}^{\dot{K}L} & = & (Q(j)^{-1})^{\dot{K}M} \Big\{
	    \left( {\delta_M}^N \, {\chi}^2 - {Q(j)_M}^N \right) {c_{\mu N}}^L 
	  - (\partial_{\mu} {j_N}^L) \, {Q(j)_M}^N \nonumber \\
   &  &   - (\partial_{\mu} j^{\dot{N}L}) \, 
	    Q(j)_{M \dot{N}} 
	    + ({\partial_{\mu}} {j_M}^L) \, {\chi}^2 \Big\}.
\end{eqnarray}
\end{proposition}

{\sl Proof.}
Using (\ref{delta2}) we can rewrite (\ref{FunInt1}) as follows:
\begin{eqnarray}
{\cal F} & = & \int \prod{\rm d}\psi \,
	 \prod{\rm d} \overline{\psi} \, \prod{\rm d}A \,
	 {\rm e}^{i \, S[A,\psi,\overline{\psi}]} \nonumber \\
  &   &  \times \, \int \prod {\rm d}j \, {\rm d}c \, 
	 \delta(j-{\cal J}) \, \delta(c-C).
\end{eqnarray}
Using Proposition 2, the first Remark after Proposition 2 and Lemma 3, we
get under the functional integral ${\cal L} = {\cal L}[j,c]$.

In a next step we integrate out the components in the diagonal blocks of the 
covector field $c_{\mu}^{ab}$. We have
\begin{eqnarray}
\lefteqn{\delta(c-C) \nonumber } \\
 & \equiv & \prod_{\mu = 1 \ldots 4 \atop a,b = 1,2,\dot{1},\dot{2}} \, 
       \delta \left( c_{\mu}^{ab} - C_{\mu}^{ab} \right) \label{deltdelt} \\
 & = & \prod_{\mu = 1 \ldots 4 \atop K,L = 1,2} \,       
       \delta \left( {c_{\mu K}}^L - {C_{\mu K}}^L \right) \,
       \delta \left( {{c_{\mu}}^{\dot{K}}}_{\dot{L}} 
       - {{C_{\mu}}^{\dot{K}}}_{\dot{L}} \right) \,
       \delta \left( c_{\mu K \dot{L}} - C_{\mu K \dot{L}} \right) \, 
       \delta \left( {c_{\mu}}^{\dot{K}L} - {C_{\mu}}^{\dot{K}L} \right). 
       \nonumber
\end{eqnarray}
Using identites (\ref{Id2c}) and (\ref{Id2d}), together with (\ref{delta4}),
we get
\begin{eqnarray*}
\lefteqn{\prod_{\mu = 1 \ldots 4 \atop K,L = 1,2} \,	   
       \delta \left( c_{\mu K \dot{L}} - C_{\mu K \dot{L}} \right) \, 
       \delta \left( {c_{\mu}}^{\dot{K}L} - {C_{\mu}}^{\dot{K}L} \right)} \\
 & = & \prod_{\mu = 1 \ldots 4 \atop K,L = 1,2} \,	 
       \delta \left( c_{\mu K \dot{L}} 
	 - (Q(j)^{-1})_{K \dot{N}} Q(j)^{\dot{N}M} C_{\mu M \dot{L}} \right) \,
       \delta \left( {c_{\mu}}^{\dot{K}L} 
	 - (Q(j)^{-1})^{\dot{K}N} Q(j)_{N \dot{M}} 
{C_{\mu}}^{\dot{M}L} \right) \\
 & = & \prod_{\mu = 1 \ldots 4 \atop K,L = 1,2} \,       
       \delta \left( c_{\mu K \dot{L}} 
- (Q(j)^{-1})_{K \dot{N}} Q({\cal J})^{\dot{N}M} C_{\mu M \dot{L}} \right) \\
 &   & \hspace*{0.85cm} \times \, \delta \left( {c_{\mu}}^{\dot{K}L} 
- (Q(j)^{-1})^{\dot{K}N} Q({\cal J})_{N \dot{M}} {C_{\mu}}^{\dot{M}L} 
\right) \\
 & = & \prod_{\mu = 1 \ldots 4 \atop K,L = 1,2} \,       
       \delta \Big( c_{\mu K \dot{L}} 
	  - (Q(j)^{-1})_{K \dot{N}} 
	    \bigg\{ \left( {\delta^{\dot{N}}}_{\dot{M}} \, {\cal X}^2 
	  - {Q({\cal J})^{\dot{N}}}_{\dot{M}} \right) 
	    {{C_{\mu}}^{\dot{M}}}_{\dot{L}}  \\
 &   &	  \hspace*{1.75cm} - (\partial_{\mu} {\cal J}_{M \dot{L}}) 
\, Q({\cal J})^{\dot{N}M} 
	  - (\partial_{\mu} {{\cal J}^{\dot{M}}}_{\dot{L}}) \, 
	    {Q({\cal J})^{\dot{N}}}_{\dot{M}} 
	  + (\partial_{\mu} {{\cal J}^{\dot{N}}}_{\dot{L}}) \, 
{\cal X}^2 \bigg\} \Big) \\
 &   & \hspace*{0.85cm} \times \, \delta \Big( {c_{\mu}}^{\dot{K}L} 
	  - (Q(j)^{-1})^{\dot{K}N} 
	    \bigg\{ \left( {\delta_N}^M \, {\cal X}^2 
	  - {Q({\cal J})_N}^M \right) {C_{\mu M}}^L \\
 &   &	  \hspace*{1.75cm} - (\partial_{\mu} {{\cal J}_M}^L) \, 
	    {Q({\cal J})_N}^M 
	  - (\partial_{\mu} {\cal J}^{\dot{M}L}) \, 
	    Q({\cal J})_{N \dot{M}} 
	  + (\partial_{\mu} {{\cal J}_N}^L) \, {\cal X}^2  \bigg\} \Big) \\
 & = & \prod_{\mu = 1 \ldots 4 \atop K,L = 1,2} \,       
       \delta \Big( c_{\mu K \dot{L}} 
	  - (Q(j)^{-1})_{K \dot{N}} \bigg\{
	    \left( {\delta^{\dot{N}}}_{\dot{M}} \, 
{\chi}^2 - {Q(j)^{\dot{N}}}_{\dot{M}} \right) 
	    {{C_{\mu}}^{\dot{M}}}_{\dot{L}} \\
 &   &	  \hspace*{1.75cm} - (\partial_{\mu} j_{M \dot{L}}) 
\, Q(j)^{\dot{N}M} 
	  - (\partial_{\mu} {j^{\dot{M}}}_{\dot{L}}) \, 
	    {Q(j)^{\dot{N}}}_{\dot{M}} 
+ (\partial_{\mu} {{j}^{\dot{N}}}_{\dot{L}}) \, {\chi}^2 \bigg\} \Big) \\
 &   & \hspace*{0.85cm} \times \, \delta \Big( {c_{\mu}}^{\dot{K}L} 
	  - (Q(j)^{-1})^{\dot{K}N} \bigg\{
	    \left( {\delta_N}^M \, {\chi}^2 - {Q(j)_N}^M \right) 
{C_{\mu M}}^L	 \\
 &   &	  \hspace*{1.75cm} - (\partial_{\mu} {j_M}^L) \, {Q(j)_N}^M 
	  - (\partial_{\mu} j^{\dot{M}L}) \, Q(j)_{N \dot{M}} 
	  + (\partial_{\mu} {{j}_N}^L) \, {\chi}^2 \bigg\} \Big) \\
 & = & \prod_{\mu = 1 \ldots 4 \atop K,L = 1,2} \,       
       \delta \Big( c_{\mu K \dot{L}} 
	  - (Q(j)^{-1})_{K \dot{N}} \bigg\{
	    \left( {\delta^{\dot{N}}}_{\dot{M}} \, 
{\chi}^2 - {Q(j)^{\dot{N}}}_{\dot{M}} \right) 
	    {{c_{\mu}}^{\dot{M}}}_{\dot{L}} \\
 &   &	  \hspace*{1.75cm} - (\partial_{\mu} j_{M \dot{L}}) \, 
Q(j)^{\dot{N}M} 
	  - (\partial_{\mu} {j^{\dot{M}}}_{\dot{L}}) \, 
	    {Q(j)^{\dot{N}}}_{\dot{M}} 
	  + (\partial_{\mu} {{j}^{\dot{N}}}_{\dot{L}}) \, 
{\chi}^2   \bigg\} \Big) \\
 &   & \hspace*{0.85cm} \times \, \delta \Big( {c_{\mu}}^{\dot{K}L} 
	  - (Q(j)^{-1})^{\dot{K}N} \bigg\{
	    \left( {\delta_N}^M \, {\chi}^2 - {Q(j)_N}^M \right)
 {c_{\mu M}}^L \\
 &   &	  \hspace*{1.75cm} - (\partial_{\mu} {j_M}^L) \, {Q(j)_N}^M 
	  - (\partial_{\mu} j^{\dot{M}L}) \, Q(j)_{N \dot{M}} 
	  + (\partial_{\mu} {{j}_N}^L) \, {\chi}^2 \bigg\} \Big).
\end{eqnarray*}
In the last step we used the first two $\delta$-functions of
(\ref{deltdelt}) to substitute ${{C_{\mu}}^{\dot{M}}}_{\dot{L}}$ and
${C_{\mu M}}^L$ by its corresponding c-number quantities.
$(Q(j)^{-1})_{K \dot{N}}$ and $(Q(j)^{-1})^{\dot{K}N}$ denote the
inverse of the $2 \times 2 \, $-matrices $Q(j)^{\dot{N}K}$ and
$Q(j)_{N\dot{K}}$, respectively, and $\chi^2$ is given by (\ref{chi}).
Now we can perform the integration over $c_{\mu K \dot{L}}$ and
${c_{\mu}}^{\dot{K}L}$, which yields
\begin{eqnarray}
{\cal F} & = & \int \prod{\rm d}\psi \,
	 \prod{\rm d} \overline{\psi} \, \prod{\rm d}A \,
	 \int \prod {\rm d}j \, \delta(j-{\cal J}) \nonumber \\
  &   &  \times \, \int \prod_{\mu = 1 \ldots 4 \atop K,L = 1,2} \,	  
{\rm d} {c_{\mu K}}^L \, {\rm d} {{c_{\mu}}^{\dot{K}}}_{\dot{L}} \, 
	 \delta({c_{\mu K}}^L - {C_{\mu K}}^L) \, 
	 \delta({{c_{\mu}}^{\dot{K}}}_{\dot{L}} 
	 - {{C_{\mu}}^{\dot{K}}}_{\dot{L}}) \, 
	 {\rm e}^{i \, S[j,c]}, 
\end{eqnarray}
where $S[j,c] \equiv S[j,{c_{\mu K}}^L,{{c_{\mu}}^{\dot{K}}}_{\dot{L}}]$.

In a next step we integrate out the gauge potential ${A_{\mu A}}^B$.
Observe, that ${A_{\mu A}}^B$ enters ${\cal F}$ only under the 
$\delta$-distributions $\delta({c_{\mu K}}^L - {C_{\mu K}}^L)$ and
$\delta({{c_{\mu}}^{\dot{K}}}_{\dot{L}} - {{C_{\mu}}^{\dot{K}}}_{\dot{L}})$. 
Using (\ref{Cmuab1}), we get
\begin{eqnarray}
{\cal F} & = & \int \prod{\rm d}\psi \, \prod{\rm d} \overline{\psi} \, 
	 \prod_{\mu = 1 \ldots 4 \atop \alpha = 1 
\ldots 8}{\rm d}A_{\mu \alpha} \,
	 \int \prod {\rm d}j \, \delta(j-{\cal J}) \nonumber \\
  &   &  \times \, \int \prod_{\mu = 1 \ldots 4 \atop \rho = 1 \ldots 8} 
	 {\rm d} c_{\mu \rho} \, \delta( c_{\mu \rho} 
	 - {{\cal Y}_{\rho}}^{\alpha} A_{\mu \alpha} 
	 - f_{\mu \rho}(\psi, \overline{\psi} \,) ) \, 
	 {\rm e}^{i \, S[j,c]}, \label{F1}
\end{eqnarray}
where
\begin{equation}
{{\cal Y}_{\rho}}^{\alpha} := 2 i g \, 
\overline{\psi_A^a} \, g^{AB} \, \psi_{bC} \,
       {(e_{\rho})_a}^b \, {(t^{\alpha})_B}^C
\end{equation}
is a non-singular $8 \times 8 \, $-matrix, and
\begin{equation}
f_{\mu \rho}(\psi, \overline{\psi}) = 
       \left( \, \overline{\psi_A^a} \, g^{AB} \, (\partial_{\mu} \psi_{bB}) 
     + \psi_{bB} \, g^{AB} \, 
(\partial_{\mu} \overline{\psi_A^a} \, ) \right) \,
     {(e_{\rho})_a}^b, 
\end{equation}
which is a function of $\psi$ and $\overline{\psi}$ only. 
Inserting 
$1 \equiv \int \prod_{\alpha = 1 \ldots 8 \atop \rho = 1 \ldots 8} 
{\rm d}{y_{\rho}}^{\alpha} \, \delta({y_{\rho}}^{\alpha} - 
{{\cal Y}_{\rho}}^{\alpha})$, 
where ${y_{\rho}}^{\alpha}$ is the c--number mate associated with 
${{\cal Y}_{\rho}}^{\alpha}$, under the functional integral we have
\begin{eqnarray*}
\delta \left( c_{\mu \rho} - {{\cal Y}_{\rho}}^{\alpha} A_{\mu \alpha} 
	- f_{\mu {\rho}}(\psi, \overline{\psi}) \right)    
  & = & \delta \left( c_{\mu {\rho}} 
	- {{\cal Y}_{\rho}}^{\beta} {y^{\sigma}}_{\beta} 
{(y^{-1})^{\alpha}}_{\sigma} \, 
	  A_{\mu \alpha} 
	- f_{\mu {\rho}}(\psi, \overline{\psi} \, ) \right) \\
  & = & \delta \left( c_{\mu {\rho}} 
	- {{\cal Y}_{\rho}}^{\beta} {{\cal Y}^{\sigma}}_{\beta} 
{(y^{-1})^{\alpha}}_{\sigma} \, 
	  A_{\mu \alpha} 
	- f_{\mu {\rho}}(\psi, \overline{\psi} \, ) \right). 
\end{eqnarray*}
Here ${(y^{-1})^{\alpha}}_{\rho}$ denotes the inverse of the 
${y^{\rho}}_{\alpha}$. A simple calculation shows, that
\begin{eqnarray*}
{{\cal Y}_{\rho}}^{\beta} {{\cal Y}^{\sigma}}_{\beta} & = & - 4 g^2 \,
       \overline{\psi_A^a} \,
       g^{AB} \, \psi_{bC} \, {(e_{\rho})_a}^b \, {(t^{\beta})_B}^C \,
       \overline{\psi_D^c} \, g^{DE} \, \psi_{dF} \, 
       {(e^{\sigma})_c}^d \, {(t_{\beta})_E}^F \\
 & = & 4 g^2 \, \left\{ \frac{2}{3} \, \overline{\psi_A^a}
       g^{AB} \, \psi_{bB} \, \overline{\psi_D^c} \, g^{DE} \, \psi_{dE} 
       -2 \, \overline{\psi_A^a} g^{AB} \, \psi_{bE} \, 
       \overline{\psi_D^c} \, g^{DE} \, \psi_{dB} \right\} \,
       {(e_{\rho})_a}^b \, {(e^{\sigma})_c}^d \\
 & = & 4 g^2 \, \left\{ \frac{2}{3} \, {{\cal J}^a}_b {{\cal J}^c}_d
       + 2 \, {{\cal J}^a}_d {{\cal J}^c}_b \right\} \,
       {(e_{\rho})_a}^b \, {(e^{\sigma})_c}^d \\
 & \equiv & {M({\cal J})_{\rho}}^{\sigma}, 
\end{eqnarray*}
where we have used the following property of the Gell--Mann--matrices:
\begin{equation}
{(t^{\beta})_B}^C \, {(t_{\beta})_E}^F
  = - \, \frac{2}{3} \, {\delta_B}^C \, {\delta_E}^F
    + 2 \, {\delta_B}^F \, {\delta_E}^C.
\end{equation}
A long, but straightforward calculation shows that
${M({\cal J})_{\rho}}^{\sigma}$ is
a non-singular $8 \times 8 \, $-matrix. Thus we obtain
\begin{eqnarray*}
\delta \left( c_{\mu {\rho}} - {{\cal Y}_{\rho}}^{\alpha} A_{\mu \alpha} 
       - f_{\mu {\rho}}(\psi, \overline{\psi} \, ) \right)    
 & = & \delta \left( c_{\mu {\rho}} 
       - {M({\cal J})_{\rho}}^{\sigma} {(y^{-1})^{\alpha}}_{\sigma} 
\, A_{\mu \alpha} 
       - f_{\mu {\rho}}(\psi, \overline{\psi} \, ) \right) \\
 & = & \delta \left( c_{\mu {\rho}} 
       - {M(j)_{\rho}}^{\sigma} {(y^{-1})^{\alpha}}_{\sigma} 
\, A_{\mu \alpha} 
       - f_{\mu {\rho}}(\psi, \overline{\psi} \, ) \right). 
\end{eqnarray*}
Now, performing the transformation
\begin{equation}
\tilde{A}_{\mu \rho} = {M(j)_{\rho}}^{\sigma} {(y^{-1})^{\alpha}}_{\sigma} \, 
		       A_{\mu \alpha},
\end{equation}
the functional integral (\ref{F1}) takes the form
\begin{eqnarray*}
{\cal F} & = & \int \prod{\rm d}\psi \, \prod{\rm d} \overline{\psi} \,
	\prod_{\mu = 1 \ldots 4 \atop \rho = 1 
\ldots 8}{\rm d}\tilde{A}_{\mu {\rho}} \,
	\int \prod {\rm d}j \, \delta(j-{\cal J}) \,
	\int \prod_{\alpha = 1 \ldots 8 \atop \rho = 1 \ldots 8} 
	{\rm d}{y_{\rho}}^{\alpha} \, 
\delta({y_{\rho}}^{\alpha} - {{\cal Y}_{\rho}}^{\alpha}) \\
  &   & \times \, \int \prod_{\mu = 1 \ldots 4 \atop \rho = 1 \ldots 8} 
	{\rm d} c_{\mu \rho} \, \delta \left( c_{\mu \rho} - 
\tilde{A}_{\mu \rho} 
	- f_{\mu \rho}(\psi, \overline{\psi} \, ) \right) \,
	\left( {\rm det} \left[ {M(j)_{\rho}}^{\sigma} 
	{(y^{-1})^{\alpha}}_{\sigma} \right] \right)^{-1} \,
	{\rm e}^{i \, S[j,c]} \\
  & = & \int \prod{\rm d}\psi \, \prod{\rm d} \overline{\psi} \, 
	\prod_{\mu = 1 \ldots 4 \atop \rho = 1 
\ldots 8}{\rm d}\tilde{A}_{\mu \rho} \,
	\int \prod {\rm d}j \, \delta(j-{\cal J}) \,
	\int \prod_{\alpha = 1 \ldots 8 \atop \rho = 1 \ldots 8} 
	{\rm d}{y_{\rho}}^{\alpha} \, 
\delta({y_{\rho}}^{\alpha} - {{\cal Y}_{\rho}}^{\alpha}) \\
  &   & \times \, 
	\int \prod_{\mu = 1 \ldots 4 \atop \rho = 1 \ldots 8} 
	{\rm d} c_{\mu \rho} \, \delta 
\left( c_{\mu \rho} - \tilde{A}_{\mu \rho} 
	- f_{\mu \rho}(\psi, \overline{\psi} \, ) \right) \,
	\frac{1}{{\rm det}[{M(j)_{\rho}}^{\sigma}]} \, 
	{\rm det} \left[ {y^{\sigma}}_{\alpha} \right] \,
	{\rm e}^{i \, S[j,c]}.
\end{eqnarray*}
Now we can trivially integrate out $\tilde{A}_{\mu \rho}$ and the
auxiliary field ${y_{\rho}}^{\alpha}$. We get
$$
{\cal F} = \int \prod{\rm d}\psi \, \prod{\rm d} \overline{\psi} \, 
        \int \prod {\rm d}j \, \delta(j-{\cal J}) \,
	\int \prod_{\mu = 1 \ldots 4 \atop \rho = 1 \ldots 8} 
	{\rm d} c_{\mu \rho} \, \frac{1}{{\rm det}[{M(j)_{\rho}}^{\sigma}]} \, 
        {\rm det} \left[ {{\cal Y}^{\sigma}}_{\alpha} \right] \,
        {\rm e}^{i \, S[j,c]}.
$$	  
It remains to calculate ${\rm det} [{{\cal Y}^{\sigma}}_{\alpha}]$.
A very long but straightforward calculation shows, that 
$({\cal J}^2)^2 \, {\rm det} \left[ {{\cal Y}^{\sigma}}_{\alpha} \right]$ is 
a nonvanishing element of maximal rank. Thus -- due to Lemma 1 -- there exists
a nonzero real number $a$ such that
\begin{equation}
({\cal J}^2)^2 \, {\rm det} \left[ {{\cal Y}^{\sigma}}_{\alpha} \right]
 = a \, ({\cal X}^2)^4.
\end{equation}
Inserting -- due to (\ref{delta4}) -- an additional factor
$\int \prod {\rm d}{\tilde j} \, \delta(j-{\tilde j})$
under the functional integral we can write
\begin{equation}
{\rm det} \left[ {{\cal Y}^{\sigma}}_{\alpha} \right] =
       \frac{(j^2)^2}{({\tilde j}^2)^2} \,
       {\rm det} \left[ {{\cal Y}^{\sigma}}_{\alpha} \right] \\
 = \frac{({\cal J}^2)^2} {({\tilde j}^2)^2} \,
       {\rm det} \left[ {{\cal Y}^{\sigma}}_{\alpha} \right] \\    
 = a \frac{({\cal X}^2)^4} {({\tilde j}^2)^2}.
\end{equation}
Now we can integrate out the auxiliary quantity $\tilde j$  and obtain
\begin{eqnarray*}
{\cal F} & = & \int \prod{\rm d}\psi \, \prod{\rm d} \overline{\psi} \, \int
	\prod {\rm d}j \, {\rm d}c \, \delta(j-{\cal J}) \,	     
	\frac{(\chi^2)^4}{{\rm det}[{M(j)_{\rho}}^{\sigma}] \, (j^2)^2} 
	{\rm e}^{i \, S[j,c]}, \\
\end{eqnarray*}
where $\chi^2$ and $j^2$ are given by (\ref{chi}) and (\ref{j}), respectively.
The number $a$ has to be absorbed in the (global) normalization factor, which
-- any way -- is omitted here.

The remaining gauge dependent fields $\psi$ and $\overline{\psi}$ occur
only in the $\delta$-distributions. To integrate them out we use the
integral representation (\ref{delta1}), i.e. we insert
\begin{eqnarray*}
\delta(j-{\cal J}) 
   & = & \int \prod {\rm d} \lambda \,
	 {\rm exp} \left[ 2 \pi i \, \lambda_{ab} \,
	 (j^{ab} - J^{ab}) \right] \\
   & = & \int \prod {\rm d} \lambda \,      
	 {\rm e}^{2 \pi i \, \lambda_{ab} j^{ab}} \, 
	 \sum_{n=0}^{12} \frac{(2 \pi i)^n}{n!} \, 
	 (- \lambda_{ab} {\cal J}^{ab})^n.
\end{eqnarray*} 
Observe that nonvanishing contributions will come from terms which are of
order 12 both in $\psi$ and $\overline{\psi}$. We get 
\begin{eqnarray*}
{\cal F} & = & \int \prod{\rm d}\psi \, \prod{\rm d}\overline{\psi} \, 
	 \int \prod {\rm d}j \, {\rm d}c \, 
	 \frac{(\chi^2)^4}{{\rm det}[{M(j)_{\rho}}^{\sigma}] \, (j^2)^2} \,
	 {\rm e}^{i \, S[j,c]} \\
   &   & \times \, \int \prod {\rm d}{\lambda} \, {\rm e}^{2 \pi i \, 
         \lambda j} 
	 \, (-1)^{12} \, \frac{(2 \pi i)^{12}}{12!} \,
	 \left( \lambda_{ab} \, J^{ab} \right)^{12} \\
   & = & \int \prod {\rm d}j \, {\rm d}c \, \frac{(2 \pi)^{12}}{12!} \,
	 \frac{(\chi^2)^4}{{\rm det}[{M(j)_{\rho}}^{\sigma}] \, (j^2)^2} \,
	 {\rm e}^{i \, S[j,c]} \,
	 \int \prod {\rm d}{\lambda} \, {\rm e}^{2 \pi i \, \lambda j} \,
	 \prod_{i=1}^{12} \, \lambda_{a_i b_i} \\
   &   & \times \, \int \prod{\rm d}\psi \, \prod{\rm d}\overline{\psi} \, 
	 \prod_{i=1}^{12} J^{a_i b_i}. 
\end{eqnarray*}
Now we can integrate out $\psi$ and $\overline{\psi}$ using equation 
(\ref{AppB4}). Replacing the factors $\lambda_{ab}$ by corresponding 
derivatives $\frac{\partial}{\partial j^{ab}}$ yields
$$
{\cal F} = \int \prod {\rm d}j \, {\rm d}c \, K[j] \, {\rm e}^{i \, S[j,c]} \,
	 \int \prod {\rm d}{\lambda} \, {\rm e}^{2 \pi i \, \lambda j},
$$
where
\begin{eqnarray*}
K[j] & = & \frac{6^4}{12!} \,
	 \frac{(\chi^2)^4}{{\rm det}[{M(j)_{\rho}}^{\sigma}] \, (j^2)^2} \,
	 \epsilon_{c_1 c_4 c_7 c_{10}} \, \epsilon_{c_2 c_5 c_8 c_{11}} \,
	 \epsilon_{c_3 c_6 c_9 c_{12}} \, \epsilon^{b_1 b_4 b_7 b_{10}} \,
	 \epsilon^{b_2 b_5 b_8 b_{11}} \, \epsilon^{b_3 b_6 b_9 b_{12}} \\
  &   &  \left\{ \times \, \prod_{r=1}^{12} \beta^{a_r c_r} \, 
	 \frac{\partial}{\partial j^{a_r b_r}} \right\} \,
	 \int \prod {\rm d}{\lambda} \, {\rm e}^{2 \pi i \, \lambda j} \, .
\end{eqnarray*} 
With
$$
\int \prod {\rm d}{\lambda} \, {\rm e}^{2 \pi i \, \lambda j} = \delta(j)
$$
we can perform the integration over $\lambda$, which finally proves the 
Proposition.
\hfill $\Box$ \vspace*{0.5cm}

We remark, that until now we considered only the ``bare'' functional
integral (\ref{FunInt1}). To calculate the vacuum expectation value for
some observable, one has to insert this observable, i.e. a gauge
invariant function ${\cal O}[A,\psi,\overline{\psi}]$, under the above
integral. With the tools given in the last sections we see, that this
function can be written down in the form ${\cal O} = {\cal
O}[C_{\mu}^{ab}, {\cal J}^{ab} \,]$. Our method of changing variables
used in the proof of Proposition 2 works in the same way for vacuum
expectation values of observables of this type, yielding an additional
factor ${\cal O}[c_{\mu}^{ab}, j^{ab}]$. This shows that, indeed, we
are dealing with a reduced theory: vacuum expectation values of baryons
(trilinear combinations of quarks) can -- in general -- not be
calculated using the above functional integral. Only certain
combinations, namely -- roughly speaking -- such, which are expressable
in terms of the $j$-field, may be treated this way. 
This is due to the fact that there exist
certain identities relating bilinear combinations of quarks and
antiquarks at one hand and trilinear combinations of quarks and their
complex conjugates on the other hand, see formula (D.7) of \cite{KR4}.

\begin{appendix}
\renewcommand{\theequation}{\Alph{section}.
\arabic{equation}}
\setcounter{equation}{0}
\section{Spinorial structures}

Since we are going to work with multilinear (and not only bilinear)
expressions in spinor fields, the ordinary matrix notation is not
sufficient for our purposes. Therefore, we will have to use a
consequent tensorial calculus in bispinor space. For those, who are not
familiar with this language we give a short review of its basic
notions. A bispinor will be represented by:
\begin{equation}\label{appendix1}
\psi^a = \left( \begin{array}{c} \phi^K \\
\varphi_{\dot K} \end{array} \right) \equiv \left(
\begin{array}{c} \phi^1 \\ \phi^2 \\ \varphi_{\dot 1} \\
\varphi_{\dot 2} \end{array} \right) \, ,
\end{equation}
where $\phi^K$ is a Weyl spinor belonging to the spinor space ${\it S}
\cong {\mathbb C}^2$, carrying the fundamental representation of 
$SL(2,{\mathbb C})$. Besides $\it S$ we have to consider the spaces
${\it S}^*$, ${\bar {\it S}}$ and ${\bar {\it S}}^*$, where $*$ denotes
the algebraic dual and bar denotes the complex conjugate. All these
spaces are isomorphic to $\it S$, but carry different representations
of $SL(2,{\mathbb C})$.  In ${\it S}^*$ acts the dual (equivalent to
the fundamental) representation and in ${\bar {\it S}}$ acts the
conjugate (not equivalent) representation of $SL(2,{\mathbb C})$. The
space $\it S$ is equipped with an $SL(2,{\mathbb C})$-invariant,
skew-symmetric bilinear form $\epsilon_{KL}$. Since it is
non-degenerate, it gives an isomorphism between $\it S$ and ${\it
S}^*$:
\begin{equation}\label{appendix2}
\it S \ni \left(\phi^K\right) \mapsto \left(\phi_L\right)
= \left(\phi^K \, \epsilon_{KL}\right) \in  {\it S}^*.
\end{equation}
There is also a canonical anti-isomorphism between
$\it S$ and ${\bar {\it S}}$ given by complex conjugation:
\begin{equation}\label{appendix3}
\it S \ni \left(\phi^K\right) \mapsto
\left({\bar \phi}^{\dot K} \right) \equiv
\left({\overline {\phi^K}} \right) \,
\in  {\bar {\it S}}.
\end{equation}
Finally, the conjugate bilinear form ${\epsilon}_{{\dot K}{\dot L}}$
gives an isomorphism between $\bar {\it S}$ and ${\bar {\it S}}^*$:
\begin{equation}\label{appendix4}
{\bar {\it S}} \ni \left(\varphi^{\dot K} \right) \mapsto
\left(\varphi_{\dot L} \right) = \left(\varphi^{\dot K}
\, \epsilon_{{\dot K}{\dot L}}\right) \in
{\bar {\it S}}^*.
\end{equation}
To summarize, we have the following commuting diagram:
\begin{equation}\label{appendix5}
\begin{array}{ccc} \it S \\ \downarrow \\ {\it S}^* \\
\end{array} \begin{array}{ccc} & \longrightarrow & \\ &&
\\ & \longrightarrow & \\ \end{array}
\begin{array}{ccc} {\bar {\it S}} \\ \downarrow \\
{\bar {\it S}}^* \\ \end{array}
\begin{array}{ccc} && \\ && \\ . \\ \end{array}
\end{equation}
Formula (\ref{appendix1}) means that a bispinor is an element of 
${\cal S} := {\it S} \times {\bar {\it S}}^*$, carrying the product of the 
fundamental and the dual to the conjugate representation of 
$SL(2,{\mathbb C})$. 
We also consider the complex conjugate bispinor
\begin{equation}\label{appendix6}
{\bar \psi}^a = \left( \begin{array}{c}
{\bar \phi}^{\dot L} \\ {\bar \varphi}_L \end{array}
\right) \equiv \left(
\begin{array}{c} {\bar \phi}^{\dot 1} \\
{\bar \phi}^{\dot 2} \\ {\bar \varphi}_1 \\
{\bar \varphi}_2 \end{array} \right) \, ,
\end{equation}
belonging to the conjugate space 
${\bar {\cal S}} = {\bar {\it S}} \times {\it S}^*$. The tensor product of
$\epsilon_{KL}$ and $- \epsilon^{{\dot K} {\dot L}}$ defines a skew symmetric 
bilinear form $\epsilon_{ab}$ on ${\cal S}$, which in turn gives an 
isomorphism 
between ${\cal S}$ and ${\cal S}^*$:
\begin{equation}\label{appendix7}
{\cal S} \ni \left( \psi^a \right) \mapsto
\left( \psi_b \right) = \left( \psi^a \, \epsilon_{ab}
\right) \in {\cal S}^* \, ,
\end{equation}
with
\begin{equation}\label{appendix8}
\epsilon_{ab} =
\left( \begin{array}{ccc} & \vline & \\ \epsilon_{KL} &
\vline & 0  \\ & \vline & \\ \hline & \vline & \\ 0 &
\vline & - \epsilon^{{\dot K}{\dot L}} \\ & \vline &
\end{array} \right) \, .
\end{equation}
We choose the minus sign in the lower block, because lowering the bispinor 
index $\uparrow a =(\uparrow K,\downarrow {\dot K})$ according to 
(\ref{appendix7}) means lowering $K$ and rising ${\dot K}$. But rising an 
index needs $-\epsilon$, because we have
\begin{equation}\label{appendix9}
\epsilon_{{\dot K}{\dot L}} \,
\epsilon^{{\dot L}{\dot M}} = - {\delta_{\dot K}}^{\dot M}
\, .
\end{equation}
The natural algebraic duality between ${\cal S}$ and
${\bar {\cal S}} = {\bar {\it S}} \times {\it S}^*
\cong {\it S}^* \times {\bar {\it S}}$ defines a
Hermitian bilinear form $\beta_{ab}$ given by
\begin{equation}\label{appendix10}
{{\bar \psi}^a_{(1)}} \, \beta_{ab} \psi^b_{(2)} =
{{\bar \phi}^{\dot K}_{(1)}} \, \varphi_{(2){\dot K}}
+ {{\bar \varphi}_{(1) K}} \, \phi_{(2)}^K \,.
\end{equation}
Thus,
\begin{equation}\label{appendix11}
\beta_{ab} =
\left( 
\begin{array}{ccc} & \vline & \\ 
0 & \vline & {\delta_{\dot K}}^{\dot L}  \\ 
& \vline & \\ 
\hline & \vline & \\ {\delta^K}_L & \vline &  0 \\ & \vline &
\end{array} \right) \, .
\end{equation}
(We stress that $a$ and $b$ are different indices: 
$\downarrow a=(\downarrow {\dot K}, \uparrow K)$ is
a conjugate index corresponding to ${\bar {\cal S}}$ and
$\downarrow b =(\downarrow  L, \uparrow {\dot L})$ 
is an index from ${\cal S}$.) The relations between spin tensors and space 
time objects are given by the Dirac $\gamma$-matrices, which we use in the 
chiral representation
\begin{equation}\label{appendix12}
\gamma^{\mu} =
\left( \begin{array}{ccc} 0 & \vline &
{\tilde \sigma}^{\mu}  \\ \hline \sigma^{\mu} & \vline &
0 \end{array} \right) \, ,
\end{equation}
with
\begin{equation}\label{appendix13}
{\tilde \sigma}^{\mu} = -\epsilon \, {\bar \sigma}^{\mu}
\, \epsilon = \left( {\bf 1}, - \sigma^k \right) \,.
\end{equation}
In index notation we have
\begin{equation}\label{appendix14}
\sigma^{\mu} = \left( {\sigma^{\mu}}_{{\dot K}L} \right)
\, ,
\end{equation}
and
\begin{equation}\label{appendix15}
{\tilde \sigma}^{\mu} =
\left( {\tilde \sigma}^{\mu M{\dot N}} \right) \, ,
\end{equation}
with
\begin{equation}\label{appendix16}
{\tilde \sigma}^{\mu M{\dot N}} = - \epsilon^{MK} \,
\overline{{\sigma^{\mu}}_{{\dot K}L}} \,
\epsilon^{{\dot L}{\dot N}} = \epsilon^{MK} \,
\epsilon^{{\dot N}{\dot L}} \,
{{\bar \sigma}^{\mu}}_{K{\dot L}}  = \epsilon^{MK} \,
\epsilon^{{\dot N}{\dot L}} \, {\sigma^{\mu}}_{{\dot L}K}
= {\sigma^{\mu}}^{{\dot N}M}.
\end{equation}
Finally, we get
\begin{equation}\label{appendix17}
({\gamma^{\mu}})^a{}_b =
\left( \begin{array}{ccc} & \vline & \\ 0 & \vline &
\sigma^{\mu {\dot L} K}  \\ & \vline & \\ \hline & \vline
& \\ {\sigma^{\mu}}_{{\dot K} L} & \vline &  0 \\ &
\vline & \end{array} \right) \, ,
\end{equation}
where $\uparrow a = (\uparrow K,\downarrow {\dot K})$ and 
$\downarrow b = (\downarrow L,\uparrow {\dot L})$.
After pulling down the first index by the help of $\epsilon_{ac}$, where
$\downarrow c = (\downarrow M,\uparrow {\dot M})$, we get
\begin{equation}\label{appendix18}
({\gamma^{\mu}})_{cb} = {(\gamma^{\mu})^a}_b \, \epsilon_{ac} =
\left( \begin{array}{ccc} & \vline & \\ 0 & \vline &
{\sigma^{\mu {\dot L}}}_M  \\ & \vline & \\ \hline &
\vline & \\ {\sigma^{\mu \dot M}}_L & \vline &  0 \\ &
\vline & \end{array} \right) \, ,
\end{equation}
which is a symmetric bilinear form, because
\begin{equation}\label{appendix19}
\psi_{(1)}^c \, {\gamma^{\mu}}_{cb} \, \psi_{(2)}^b =
\phi_{(1)}^M \, {\sigma^{\mu {\dot L}}}_M \,
\varphi_{(2){\dot L}} + \varphi_{(1){\dot M}} \,
{\sigma^{\mu {\dot M}}}_L \, \phi_{(2)}^L \, .
\end{equation}
The complex conjugate quantity is given by
\begin{equation}\label{appendix20}
\overline{({\gamma^{\mu}})_{cb}} =
\left( \begin{array}{ccc} & \vline & \\ 0 & \vline &
{{\sigma^{\mu}}_{\dot M}}^L  \\ & \vline & \\ \hline &
\vline & \\ {{\sigma^{\mu}}_{\dot L}}^M  & \vline &  0 \\
& \vline & \end{array} \right) \, .
\end{equation}
We also use the following spin tensor
\begin{equation}\label{appendix21}
\beta_{abcd} := \frac{1}{2}
\overline{({\gamma^{\mu}})_{ab}} \,
({\gamma_{\mu}})_{cd} \, .
\end{equation}
Obviously, this tensor is symmetric in the first and in the second pair of 
indices seperately. We see from (\ref{appendix19}) and (\ref{appendix20}) 
that 
it vanishes, whenever $a$ and $b$ or $c$ and $d$ are of the same type (both 
dotted or both undotted). Thus, the only nonvanishing components are
\begin{equation}\label{appendix22}
\beta_{\dot K}{}^{L{\dot M}}{}{}_N 
  = \beta_{\dot K}{}^{L}{}_N{}^{\dot M} 
  = \beta^{L}{}_{\dot K}{}^{\dot M}{}_N 
  = \beta^{L}{}_{\dot K}{}_N{}^{\dot M} 
  = - {\delta^{\dot M}}_{\dot K} \, {\delta^L}_N \, .
\end{equation}
Finally, observe that second rank spin tensors have a natural block structure.
In particular, for the invariants considered in this paper we get:
\begin{equation}\label{appendix23}
{\cal J}^{ab} = \left( \begin{array}{ccc} 
& \vline & \\ 
{\cal J}^{\dot{K}L} & \vline & {{\cal J}^{\dot{K}}}_{\dot{L}} \\ 
& \vline & \\ 
\hline & \vline & \\ 
{{\cal J}_K}^L & \vline & {\cal J}_{K\dot{L}} \\ 
& \vline & \end{array} \right) \, ,
\end{equation}
where ${\cal J}^{\dot{K}L}$, ${\cal J}_{K\dot{L}}$, 
${{\cal J}^{\dot{K}}}_{\dot{L}}$ and ${{\cal J}_K}^L$ are Hermitean 
$2 \times 2 \, $-matrices. Analogously,
\begin{equation}\label{appendix24}
C_{\mu}^{ab} = \left( \begin{array}{ccc} 
& \vline & \\ 
C_{\mu}^{\dot{K}L} & \vline & {{C_{\mu}}^{\dot{K}}}_{\dot{L}} \\ 
& \vline & \\ 
\hline & \vline & \\ 
{C_{\mu K}}^L & \vline & C_{\mu K \dot{L}} \\ 
& \vline & \end{array} \right) \, . 
\end{equation}

\setcounter{equation}{0}
\section{Calculation of $({\cal X}^2)^4$}

Obviously $({\cal X}^2)^4$ is an element of maximal rank in the
Grass\-mann-algebra, that is
\begin{equation}
\label{AppB1}
({\cal X}^2)^4 = c \, \prod \psi \, \prod {\bar \psi}.
\end{equation}
To prove that it is nonzero it remains to calculate the number $c \in
\mathbb C$ and to
show, that $c \neq 0$. From (\ref{AppB1}) it follows by integration that
\begin{equation}
\label{AppB2}
c = \int ({\cal X}^2)^4 \, \prod {\rm d}\psi \, \prod {\rm d}{\bar \psi}.
\end{equation}
Using the definition of ${\cal X}^2$ we have
$$
({\cal X}^2)^4 = \prod_{r=1}^4 \, 4 \, 
	 ({\cal J}^{a_r d_r} {\cal J}^{b_r e_r} {\cal J}^{c_r f_r} +
	 2 \, {\cal J}^{a_r f_r} {\cal J}^{b_r d_r} {\cal J}^{c_r e_r} )
	 \beta_{b_r c_r e_r f_r} \beta_{a_r d_r}.
$$
Inserting this into (\ref{AppB2}) we can decompose $c$ into a sum
\begin{equation}
\label{AppB3}
c = c_1 + c_2 + c_3 + c_4 + c_5,
\end{equation}
with
\begin{eqnarray}
c_1 & = & 4^4 \, \int \prod_{r=1}^4 \, 
	  {\cal J}^{a_r d_r} {\cal J}^{b_r e_r} {\cal J}^{c_r f_r}
	  \beta_{b_r c_r e_r f_r} \, \beta_{a_r d_r} \,
	  \prod {\rm d}\psi \, \prod {\rm d}{\bar \psi} \nonumber \\
    & = & \frac{1}{(4!)^3} \, 6^4 \, (4!)^3 \, 4^4 \, \prod_{r=1}^4 
	  ( \beta_{b_r c_r e_r f_r} \, \beta_{a_r d_r} \,
	  \beta^{a_r i_r} \, \beta^{b_r j_r} \, \beta^{c_r k_r} ) \nonumber \\
    &   & \times \, \epsilon_{i_1 i_4 j_3 k_2} \, \epsilon_{i_2 j_1 j_4 k_3} \,
	  \epsilon_{i_3 j_2 k_1 k_4} \, \epsilon^{d_1 d_4 e_3 f_2} \,
	  \epsilon^{d_2 e_1 e_4 f_3} \, \epsilon^{d_3 e_2 f_1 f_4} \nonumber \\
    & = & 6^4 \, 4^4 \, U_{e_1 f_1 d_1}^{i_1 j_1 k_1} \,
	  U_{e_2 f_2 d_2}^{i_2 j_2 k_2} \, U_{e_3 f_3 d_3}^{i_3 j_3 k_3} \,
	  U_{e_4 f_4 d_4}^{i_4 j_4 k_4} \nonumber \\
    &   & \times \, \epsilon_{i_1 i_4 j_3 k_2} \, \epsilon_{i_2 j_1 j_4 k_3} \,
	  \epsilon_{i_3 j_2 k_1 k_4} \, \epsilon^{d_1 d_4 e_3 f_2} \,
	  \epsilon^{d_2 e_1 e_4 f_3} \, \epsilon^{d_3 e_2 f_1 f_4}, 
          \label{c1} \\
c_2 & = & 2^4 \, 4^4 \, \int \prod_{r=1}^4 \, 
	  {\cal J}^{a_r f_r} {\cal J}^{b_r d_r} {\cal J}^{c_r e_r}
	  \beta_{b_r c_r e_r f_r} \, \beta_{a_r d_r} \,
	  \prod {\rm d}\psi \, \prod {\rm d}{\bar \psi} \nonumber \\
    & = & 6^4 \, 2^4 \, 4^4 \, U_{e_1 f_1 d_1}^{i_1 j_1 k_1} \,
	  U_{e_2 f_2 d_2}^{i_2 j_2 k_2} \, U_{e_3 f_3 d_3}^{i_3 j_3 k_3} \,
	  U_{e_4 f_4 d_4}^{i_4 j_4 k_4} \nonumber \\
    &   & \times \, \epsilon_{i_1 i_4 j_3 k_2} \, \epsilon_{i_2 j_1 j_4 k_3} \,
	  \epsilon_{i_3 j_2 k_1 k_4} \, \epsilon^{d_1 d_4 e_3 f_2} \,
	  \epsilon^{d_2 e_1 e_4 f_3} \, \epsilon^{d_3 e_2 f_1 f_4}, 
          \label{c2} \\
    & = & 2^4 \, c_1, \nonumber \\
c_3 & = & 2^3 \, 4^4 \, \int \prod_{r=1}^3 \, 
	  {\cal J}^{a_r d_r} {\cal J}^{b_r e_r} {\cal J}^{c_r f_r}
	  \beta_{b_r c_r e_r f_r} \, \beta_{a_r d_r} \nonumber \\
    &   & \times \, {\cal J}^{a_4 f_4} {\cal J}^{b_4 d_4} {\cal J}^{c_4 e_4} \,
	  \beta_{b_4 c_4 e_4 f_4} \, \beta_{a_4 d_4} \,
	  \prod {\rm d}\psi \, \prod {\rm d}{\bar \psi} \nonumber \\
    & = & 6^4 \, 2^3 \, 4^4 \, U_{e_1 f_1 d_1}^{i_1 j_1 k_1} \,
	  U_{e_2 f_2 d_2}^{i_2 j_2 k_2} \, U_{e_3 f_3 d_3}^{i_3 j_3 k_3} \,
	  U_{e_4 f_4 d_4}^{i_4 j_4 k_4} \nonumber \\
    &   & \times \, \epsilon_{i_1 i_4 j_3 k_2} \, \epsilon_{i_2 j_1 j_4 k_3} \,
	  \epsilon_{i_3 j_2 k_1 k_4} \, \epsilon^{d_1 f_4 e_3 f_2} \,
	  \epsilon^{d_2 e_1 d_4 f_3} \, \epsilon^{d_3 e_2 f_1 e_4}, 
          \label{c3} \\
c_4 & = & 2^2 \, 4^4 \, 6 \, \int \prod_{r=1}^2 \, 
	  {\cal J}^{a_r d_r} {\cal J}^{b_r e_r} {\cal J}^{c_r f_r}
	  \beta_{b_r c_r e_r f_r} \, \beta_{a_r d_r} \nonumber \\
    &   & \times \int \prod_{s=1}^2 \, 
	  {\cal J}^{a_s f_s} {\cal J}^{b_s d_s} {\cal J}^{c_s e_s}
	  \beta_{b_s c_s e_s f_s} \, \beta_{a_s d_s} \,
	  \prod {\rm d}\psi \, \prod {\rm d}{\bar \psi} \nonumber \\
    & = & 6^4 \, 2^2 \, 4^4 \, 6 \, U_{e_1 f_1 d_1}^{i_1 j_1 k_1} \,
	  U_{e_2 f_2 d_2}^{i_2 j_2 k_2} \, U_{e_3 f_3 d_3}^{i_3 j_3 k_3} \,
	  U_{e_4 f_4 d_4}^{i_4 j_4 k_4} \nonumber \\
    &   & \times \, \epsilon_{i_1 i_4 j_3 k_2} \, \epsilon_{i_2 j_1 j_4 k_3} \,
	  \epsilon_{i_3 j_2 k_1 k_4} \, \epsilon^{d_1 f_4 d_3 f_2} \,
	  \epsilon^{d_2 e_1 d_4 e_3} \, \epsilon^{f_3 e_2 f_1 e_4}, 
          \label{c4} 
	  \\
c_5 & = & 2^5 \, 4^4 \, \int \prod_{r=1}^3 \, 
	  {\cal J}^{a_r f_r} {\cal J}^{b_r d_r} {\cal J}^{c_r e_r}
	  \beta_{b_r c_r e_r f_r} \, \beta_{a_r d_r} \nonumber \\
    &   & \times \, {\cal J}^{a_4 d_4} {\cal J}^{b_4 e_4} {\cal J}^{c_4 f_4}
	  \beta_{b_4 c_4 e_4 f_4} \, \beta_{a_4 d_4} \,
	  \prod {\rm d}\psi \, \prod {\rm d}{\bar \psi} \nonumber \\
    & = & 6^4 \, 2^5 \, 4^4 \, U_{e_1 f_1 d_1}^{i_1 j_1 k_1} \,
	  U_{e_2 f_2 d_2}^{i_2 j_2 k_2} \, U_{e_3 f_3 d_3}^{i_3 j_3 k_3} \,
	  U_{e_4 f_4 d_4}^{i_4 j_4 k_4} \nonumber \\
    &   & \times \, \epsilon_{i_1 i_4 j_3 k_2} \, \epsilon_{i_2 j_1 j_4 k_3} \,
	  \epsilon_{i_3 j_2 k_1 k_4} \, \epsilon^{f_1 d_4 d_3 e_2} \,
	  \epsilon^{f_2 d_1 e_4 e_3} \, \epsilon^{f_3 d_2 e_1 f_4}, 
          \label{c5} 
\end{eqnarray}
where 
$$
U_{efd}^{ijk} := \beta_{bcef} \, \beta_{ad} \, \beta^{ai} \, 
		 \beta^{bj} \, \beta^{ck}.
$$
Here we made use of the following equation
\begin{eqnarray}
\label{AppB4}
\int \prod{\rm d}\psi \,
\prod{\rm d}\overline{\psi} \, \,
\prod_{r=1}^{12} {\cal J}^{c_r d_r} 
  & = & \frac{1}{(4!)^3} \, \prod_{r=1}^{12}
	\beta^{c_r l_r} \sum_{\sigma} \epsilon_
	{l_{\sigma_1} l_{\sigma_2} l_{\sigma_3} l_{\sigma_4}}\,
	\epsilon_{l_{\sigma_5} \, ... \,  l_{\sigma_8}}\,
	\epsilon_{l_{\sigma_9} \, ... \,
	l_{\sigma_{12}}}\,\nonumber \\
  &   & \times \,\, \epsilon^
	{d_{\sigma_1} d_{\sigma_2} d_{\sigma_3} d_{\sigma_4}}\,
	\epsilon^{d_{\sigma_5} \, ... \,  d_{\sigma_8}}\,
	\epsilon^{d_{\sigma_9} \, ... \, d_{\sigma_{12}}},
\end{eqnarray}
where the sum is taken over all permutations. To prove this formula we 
observe, that both sides do not vanish, iff every spinor index occurs exactly 
three times in the multi-indices $(c_r)$ and $(d_r)$. Furthermore, both sides 
are symmetric with respect to every simultaneous transposition of two pairs of 
indices, e.g. $(c_r,d_r)$ and $(c_s,d_s)$. Therefore, the indices $c_r$ may be 
ordered on both sides. After having done this, the above formula can be checked 
by inspection.

To prove (\ref{c1}) -- (\ref{c5}) we note 
that only such terms give a nonvanishing contribution, for which all indices 
within every $\epsilon$-tensor are different. The number of such permutations 
is $6^4 \, (4!)^3$ and it is easy to see that all of them give the same 
contribution. Therefore, we can replace the sum over all permutations by a 
concrete representation multiplied by the number $6^4 \, (4!)^3$.

The next step is to perform the sum over all indices in (\ref{c1}), 
(\ref{c2}), (\ref{c3}), (\ref{c4}) and (\ref{c5}). A lengthy but simple 
tensorial calculation gives
\begin{eqnarray*}
c_1 & = & 4^4 \, 6^4 \, 2^7 \, 3^2             = 2^{19} \, 3^6,  \\
c_2 & = & 4^4 \, 6^4 \, 2^4 \, 2^7 \, 3^2      = 2^{23} \, 3^6,  \\
c_3 & = & 4^4 \, 6^4 \, 2^3 \, 2^4 \, 3^3      = 2^{19} \, 3^7,  \\
c_4 & = & 4^4 \, 6^4 \, 2^3 \, 2^3 \, 3^3 \, 5 = 2^{18} \, 3^7 \, 5, \\
c_5 & = & 4^4 \, 6^4 \, 2^5 \, 2^4 \, 3^3 = 2^{21} \, 3^7.
\end{eqnarray*}
Taking the sum we get
$$
c = 2^{18} \, 3^6 \, 79.
$$
\hfill $\Box$ \vspace*{0.5cm}

\end{appendix}



\begin{thebibliography}{99}
\bibitem{KR1}
J.Kijowski and G.Rudolph, Lett. Math. Phys. {\bf 29}, 103 (1993)
\bibitem{KRR}
J.Kijowski, G.Rudolph and M.Rudolph, Lett. Math. Phys. {\bf 33}, 139 (1995)
\bibitem{FP}
L.D.Faddeev and V.N.Popov, Phys. Lett. {\bf B 25}, 30 (1967)
\bibitem{KRT}
J.Kijowski, G.Rudolph and A.Thielmann, {\it ``The algebra of observables and 
charge superselection sectors of QED on the lattice''} (in preparation)
\bibitem{G}
V.N.Gribov, Nucl. Phys. {\bf B139}, 1 (1978)
\bibitem{S}
I.M.Singer, Comm. Math. Phys. {\bf 60}, 7 (1978)
\bibitem{MVB}
P.K.Mitter and C.M.Viallet, Comm. Math. Phys. {\bf 79}, 457 (1981)\,, \\
O.Babelon and C.M.Viallet, Phys. Lett. {\bf 85B}, 246 (1979)\,, \\
W.Kondracki and J.Rogulski, prepr. 62/83/160, Inst. of Math. PAN, 
Warsaw (1983) \,,\\
P.K.Mitter, Cargese Lectures 1979, in G.`t Hooft et.al. (eds.), 
{\it Recent Development in Gauge Theories}, Plenum Press, New York 1980
\bibitem{Z}
D.Zwanziger, Nucl. Phys. {\bf B209}, 336 (1982), {\bf B323}, 513 (1989), 
{\bf B345}, 461 (1990)
\bibitem{KR2}
J.Kijowski and G.Rudolph, Nucl. Phys. {\bf B325}, 211 (1989)
\bibitem{R}
G.Rudolph, Lett. Math. Phys. {\bf 16}, 27 (1988)
\bibitem{KR3}
J.Kijowski and G.Rudolph, Phys. Rev. {\bf D31}, 856 (1985)\,,\\
G.Rudolph, Annalen der Physik, 7.Folge, Bd. 47, 2/3, 211 (1990)
\bibitem{KR4}
J.Kijowski and G.Rudolph, {\it ``One-Flavour Chromodynamics in Terms of Gauge
Invariant Quantities''}, Leipzig University preprint Nr. 1/93 (1993)
\bibitem{Ber1}
F.A.Berezin, {\it Metod vtori${\check c}$novo
kvantovania}, Nauka, Moskva 1986
\bibitem{Ber2}
F.A.Berezin, {\it The Method of Second Quantization}, Academic Press,
New York and London 1966
\bibitem{Ber3}
F.A.Berezin, {\it Introduction to Superanalysis}, D. Reidel Publ.
Comp., Dordrecht 1987
\bibitem{L}
D.Leites (ed), Seminar on Supermanifolds, No. 30, 31, Reports of Stockholm
University, 1988-13, 1988-14
\bibitem{T}
F. Treves, {\it Topological Vector Spaces, Distributions and Kernels}, 
Academic Press, Pure and Applied Math. 25, New York 1967
\end{thebibliography}
\end{document}